\documentclass[usenatbib, usedcolumn, twocolumn]{mn2e}
\usepackage{epsfig}
\usepackage{amsmath}
\usepackage{lscape}
\usepackage{longtable}
\usepackage{anysize}
\usepackage{color}
\usepackage{ulem}
\def\beq{\begin{equation}}
\def\eeq{\end{equation}}
\def\bey{\begin{eqnarray}}
\def\eey{\end{eqnarray}}

\def\Msun{\,{\rm M_\odot}}
\def\Msunh{\, h^{-1}{\rm M_\odot}}
\def\Msunhh{\, h^{-2}{\rm M_\odot}}

\def\gs{\mathrel{\raise1.16pt\hbox{$>$}\kern-7.0pt
\lower3.06pt\hbox{{$\scriptstyle \sim$}}}}
\def\ls{\mathrel{\raise1.16pt\hbox{$<$}\kern-7.0pt
\lower3.06pt\hbox{{$\scriptstyle \sim$}}}}
\def\gtsima{$\; \buildrel > \over \sim \;$}
\def\ltsima{$\; \buildrel < \over \sim \;$}
\def\prosima{$\; \buildrel \propto \over \sim \;$}
\def\gsim{\lower.5ex\hbox{\gtsima}}
\def\lsim{\lower.5ex\hbox{\ltsima}}
\def\simgt{\lower.5ex\hbox{\gtsima}}
\def\simlt{\lower.5ex\hbox{\ltsima}}
\def\simpr{\lower.5ex\hbox{\prosima}}
\def\la{\lsim}
\def\ga{\gsim}

\begin{document}
%

\title[Empirical model of star formation history]
      {An Empirical Model for the Star Formation History in Dark Matter Halos}

\author[Zhankui Lu et al.]
       {\parbox[t]{\textwidth}{
        Zhankui Lu$^{1}$\thanks{E-mail: lv@astro.umass.edu},
        H.J. Mo$^{1}$,
        Yu Lu$^{2}$,
        Neal Katz$^{1}$,
        Martin D. Weinberg$^{1}$,
        Frank C. van den Bosch$^{3}$,
        Xiaohu Yang$^{4,5}$}\\
           \vspace*{3pt} \\
$^1$Department of Astronomy, University of Massachusetts, Amherst MA 01003-9305, USA\\
$^2$Kavli Institute for Particle Astrophysics and Cosmology, Stanford, CA 94309, USA\\
$^3$Astronomy Department, Yale University, P.O. Box 208101, New Haven, CT 06520-8101, USA\\
$^4$Center for Astronomy and Astrophysics, Shanghai Jiao Tong University,
     Shanghai 200240, China\\
$^5$Key Laboratory for Research in Galaxies and Cosmology, Shanghai Astronomical 
    Observatory, \\ \, Nandan Road 80, Shanghai 200030, China}


\date{}
\pagerange{\pageref{firstpage}--\pageref{lastpage}}
\pubyear{2013}

\maketitle 

\label{firstpage}


\begin{abstract}
  We develop an empirical approach to infer the star formation 
  rate in dark matter halos from the galaxy stellar mass function 
  (SMF) at different redshifts and the local cluster galaxy 
  luminosity function (CGLF), which has a steeper faint end
  relative to the SMF of local galaxies. As satellites are typically 
  old galaxies which have been accreted earlier, this feature
  can cast important constraint on the formation of low-mass galaxies 
  at high-redshift. 
  The evolution of the SMFs suggests the star formation in high mass 
  halos ($>10^{12}\Msunh$) has to be boosted at high redshift beyond 
  what is expected from a simple scaling of the dynamical time. 
  The faint end of the CGLF implies a characteristic redshift
  $z_c \approx 2$ above which the star formation rate in low
  mass halos with masses $< 10^{11}\Msunh$ must be enhanced relative 
  to that at lower $z$. This is not directly expected from the 
  standard stellar feedback models. Also, this enhancement leads 
  to some interesting predictions, for instance, a significant 
  old stellar population in present-day dwarf galaxies with 
  $M_{\star} \le 10^{8}\Msunhh$ and steep slopes of high redshift 
  stellar mass and star formation rate functions.
\end{abstract}




\section{Introduction}
With the advent of multi-wave band deep surveys from the Hubble Space
Telescope, galaxy stellar mass (luminosity) functions (SMF or LF) 
can now be determined up to a redshift of $z\approx 8$ \citep[e.g.][]{Bradley12}.  
These observations have revealed a number of interesting properties 
about the galaxy population. The amplitude of the SMF increases
as the universe evolves from high to low redshift, while the 
characteristic stellar mass does not change significantly 
\citep[e.g.][]{PG08, Marchesini09, Marchesini12, Santini12}. 
This suggests that galaxies with masses larger than the characteristic mass
may have formed as early as $z=3$, while galaxies of lower masses
continue to grow in mass and/or in number density all the way to the
present epoch. 
For local galaxies, large redshift surveys have now made it possible to
determine the SMF of galaxies down to $10^7\Msun$
\citep[e.g.][]{Baldry12}. 
Such surveys have revealed that the Schechter function may not be
sufficient to describe
the observed luminosity function of galaxies, especially for galaxies
in massive clusters. Instead there seems to be a marked upturn at the
faint end ($M_{r}-5\log_{10} h > -17$) of the cluster galaxy luminosity 
function (CGLF) \citep{Popesso06, Milne07, Jenkins07, Barkhouse07, Banados10, Wegner11}.  
This feature has also been found for galaxies in the general field regardless 
of their environments, which are usually referred to as field galaxies 
\citep{Blanton05, Baldry12, Drory09, Pozzetti10, Loveday12}, although the upturn
appears shallower than that for cluster galaxies.

Theoretically, the standard $\Lambda$CDM model assumes that galaxies
form and evolve in dark matter halos \citep[see][for an overview]{Mo10}.  
However, the observed redshift evolution of the
galaxy population, in which massive galaxies form earlier, appears at
odds with the simplest expectations from the hierarchical nature of
dark matter halo formation in the $\Lambda$CDM scenario, where small
halos form first and subsequently merge to form larger ones.
Furthermore, the halo mass function predicted by the standard
$\Lambda$CDM model has a characteristic shape very different from that
of the galaxy SMF.  Although it can also be
described by a Schechter-like function, the halo mass function has a
much steeper slope at the low-mass end, and an exponential cutoff at a
much larger mass scale.  Thus, if star formation was equally efficient
in halos of different masses, the $\Lambda$CDM model predicts too many
low-mass (\citealt{Klypin99, Moore99}) and high-mass galaxies.  
Finally, the existence of a stronger faint end upturn in 
galaxy clusters is not expected from simple predictions
of the $\Lambda$CDM model. 
In fact the slope of the maximum circular velocity function of 
subhalos (substructure within distinct halos) of massive clusters   
is quite similar to that of distinct halos \citep{Klypin11}.

The apparent tension between the observations and the standard 
$\Lambda$CDM model indicates the complexity of the baryonic physics
that regulates the galaxy evolution. The effects of the baryonic physics can be 
studied directly using hydrodynamic simulations or semi-analytic models
(hereafter SAMs). Hydrodynamic simulations find
two basic modes by which baryonic matter is accreted into galaxies,
cold mode and hot mode, depending on the mass of the host halo (\citealt{Keres05, Keres09}). 
Other processes, especially those concerning star formation and feedback,
which determine how efficiently cold gas turns into stars, are still
beyond the capability of current computational resources to model and
uncertain subgrid prescriptions are used to model them.  In a
SAM \citep[e.g.,][]{Kauffmann93, Somerville99, Cole00,
Croton06, Kang05, Guo11, Lu11}, all the relevant processes (e.g. cooling,
star formation, feedback, etc) are modelled using parametrised
prescriptions either derived from analytic models or calibrated with
numerical simulations.  Although SAMs have produced useful predictions
about the galaxy population, many of the physical processes involved
are still poorly understood at present and some uncertain assumptions
have to be made about a number of model ingredients, such as the
efficiency of star formation and feedback.

In recent years, much effort has been made to establish the
statistical connection between galaxies and dark matter halos via the
conditional luminosity function (CLF) \citep[e.g.][]{Yang03,
vdBosch03} or the halo occupation distribution (HOD)
\citep[e.g.][]{Jing98, Peacock00, Seljak00, Scoccimarro01, Berlind02}. 
Such empirically established galaxy-dark matter halo connections 
describe how galaxies with different properties occupy halos of different masses and, therefore,
provide important insights into how galaxies form and evolve in dark
matter halos \citep{Mo10}.  With data obtained from deep,
multi-wavelength surveys, attempts have been made to establish the
relation between galaxies and their dark matter halos out to high $z$
using Abundance Matching (AM). This technique links galaxies of a given
luminosity or stellar mass to dark matter halos of a given mass by
matching the observed abundance of the galaxies to the halo abundance
obtained from the halo mass function, typically also accounting for subhalos. 
This approach was first used by \citet{Mo96} and
\citet{Mo99} to model the number density and clustering of Lyman-break
galaxies.  More recently, several studies have used this abundance
matching technique to probe the galaxy-dark matter connection out to
$z \approx 5$ \citep[e.g.][]{Vale04, Conroy06, Shankar06, Conroy09,
Moster10, Guo10, Behroozi10, Yang12, Bethermin12, Bethermin13}. 
One can infer the average star formation rate (SFR) for halos of different 
masses at different redshifts from the stellar mass-halo mass relation
\citep[e.g.][]{Behroozi12, Moster13, Yang13, Wang13}.

How does one learn about galaxy formation within the $\Lambda$CDM paradigm?
The usual approach using cosmological hydrodynamic simulations or SAMs 
is to make {\it ab initio} models of galaxy formation including all the 
physical processes that one thinks are important. One then makes predictions 
from these models and compares them with observations.
If the model does not compare well with the observations then one changes
the model, typically either by changing the parametrisations of the previously
included physical processes or by adding new physical processes.
In this paper we make a first step at a qualitatively different approach from
most past work. We attempt to ask in general terms another question. 
What do the observations require of the galaxy formation model? 
In other words, we attempt to let the data speak for themselves in a way that 
is as independent as possible of any model assumptions.  
In fact we only make three main assumptions:
that we live in a $\Lambda$ dominated cold dark matter Universe from which
we extract dark matter halo merger trees, that the SFR of
the central galaxies in such halos depends only on the halo mass and the 
redshift, and that when a galaxy becomes a satellite its star formation is quenched
exponentially and it can eventually merge with the central galaxy on a
dynamical friction timescale. 
Such simplifications allow us to explore a broad range of hypotheses 
without being restricted by our poor understanding of baryonic physics, 
such as cooling, star formation, and feedback. 
We put our model on a firm statistical footing using Bayesian inference.
We start with a very simple model to describe the SFR of central galaxies 
and only increase its complexity if the data requires it,
assessed using Bayes ratios.
We build up our model in a stepwise manner, increasing the
complexity as we add more constraining data, instead of using
a single complicated model, so that we  can clearly see how each 
SFR model is constrained by the different observations and to see
whether or not a more complex model is required.

This paper is organised as follows.  The generic form of our empirical
model is described in \S2. In \S3 the observational constraints,
including the SMF at different redshifts and the CGLF, are described 
together with the method we use to constrain the model parameters.  
In \S4, we show how we increase the complexity of our model in a series of
steps.  This results in a series of nested model families
that can reproduce more and more of the observational constraints:
Model I is able to match the SMF of local galaxies, 
Model II reproduces the evolution of the SMFs, and Model III is
the minimum model that can also explain the CGLF.
In \S5, we present the detailed predictions of Model II, 
and Model III for the stellar mass--halo mass relation, SFR--halo accretion 
rate relation, the cosmic star formation rate
density, the star formation rate function, the specific star formation
rate as a function of stellar mass, and the conditional stellar mass
function.
Finally, we discuss and summarise our results in \S6.

Throughout the paper, we use a $\Lambda$CDM cosmology with
$\Omega_{\rm m,0}=0.26$, $\Omega_{\Lambda}=0.74$, $\Omega_{\rm B}
=0.044$, $h=0.71$, $n=0.96$ and $\sigma_{8}=0.79$. This set of
parameters is consistent with the WMAP5 data (\citealt{Dunkley09,
Komatsu09}). 

\section{The Empirical Model}

\subsection{Dark halo merger histories}

Our empirical model is built upon dark halo merger histories generated 
using the algorithm developed by \citet{Parkinson08}. 
The algorithm is based on the Extended Press-Schechter (EPS) formalism 
and it is tuned to agree with the conditional mass functions
of merger trees \citep{Cole08} constructed from the Millennium Simulation 
\citep[MS,][]{Springel05}. 
As shown by \citet{Jiang13}, this algorithm is in good agreement
with simulations in terms of many other properties, such as
mass assembly history, merger rate and unevolved subhalo mass function.
For this work, the merger trees span a redshift range $0\le z\le 15$
with $100$ snapshots evenly distributed in $\ln(1+z)$ space.
The mass resolution is $2\times10^9\Msunh$.
The sets of merger trees used to predict the observational
constraints are described in \S3 in more detail.


\subsection{Star formation in central galaxies}

We assume that the SFR of the central galaxy of a halo
at a given redshift $z$ is completely determined by the virial mass of
the host halo, $M_{\rm vir} (z)$, and $z$. Then the SFR can be written as
\begin{equation}\label{SFR_general01}
{\rm SFR}\equiv {\dot M}_\star = {\dot M}_\star \left[ M_{\rm vir}(z), z \right] \,.
\end{equation}
The above equation describes an average among halos of a given mass at
a given $z$. It ignores any variations in the SFR that owe to
variations in the formation histories of individual halos of a given
mass and any large-scale environmental effects.
Note though  that the $\dot{M}_\star[M_{\rm vir}(z)]$
can still result in halos of the same $M_{\rm vir}(z)$ 
having different $M_\star$ simply because of scatter in
the halo assembly histories.

Guided by the observational demand that the star formation efficiency
must be suppressed in both low and high mass halos \citep{Yang03}, we
assume the following form for the dependence of the star formation
rate on redshift and halo mass:
\begin{equation}\label{eqn_general02}
  {\dot M}_\star =
    {\cal E} \frac{f_B M_{\rm vir}}{\tau_0} \left( 1+z \right)^{\kappa}
      (X+1)^{\alpha} 
    \left(\frac{X+\mathcal{R}}{X+1}\right)^{\beta} 
    \left(\frac{X}{X+\mathcal{R}}  \right)^{\gamma} 
    \,,
\end{equation}
where ${\cal E}$ is an overall efficiency; $f_B$ is the cosmic baryonic
mass fraction; $\tau_0$ is a dynamic timescale of the halos at the
present day, set to be $\tau_0\equiv 1/(10 H_0)$; and $\kappa$ is
fixed to be ${3/2}$ so that $\tau_0/(1+z)^{3/2}$ is roughly the
dynamical timescale at redshift $z$.  We define the quantity $X$ to be
$X \equiv M_{\rm vir} / M_{\rm c}$, where $M_{\rm c}$ is a
characteristic mass and $\mathcal{R}$ is a positive number that is
smaller than $1$.  Hence, in our model the SFR of a
galaxy depends on its dark matter halo mass through a piecewise power
law, with $\alpha$, $\beta$, and $\gamma$ being the three power
indices in the three different mass ranges separated by the two
characteristic masses $M_{\rm c}$ and $\mathcal{R}M_{\rm c}$:
\begin{equation}
 {\dot M}_{\star} \propto \frac{M_{\rm vir}}{\tau_0}
 \begin{cases}
   M_{\rm vir}^{\alpha} \, & \text{if} ~ M_{\rm vir} \gg M_{\rm c}                     \\
   M_{\rm vir}^{\beta}  \, & \text{if} ~ M_{\rm c} > M_{\rm vir} > \mathcal{R}M_{\rm c} \\
   M_{\rm vir}^{\gamma} \, & \text{if} ~ M_{\rm vir} \ll \mathcal{R} M_{\rm c}          \,.
 \end{cases}
\end{equation}
The simplest model is the one where all the model parameters, ${\cal
  E}$, $\alpha$, $\beta$, $\gamma$, $M_{\rm c}$ and $\mathcal{R}$ are
redshift-independent. In what follows, we will make more parameters
redshift-dependent whenever the observational data demands it.

\subsection{Star formation in satellite galaxies}

For satellites, the SFR has to be modelled differently. 
As a dark matter halo gets accreted by a larger one, it
becomes a subhalo and experiences environmental effects such as tidal
stripping, galaxy harassment \citep{Moore96}, and tidal disruption.
The satellite galaxy associated with the subhalo may also be affected
as it orbits in the host halo.  For example, the diffuse gas initially
in the subhalo and the cold gas disk may get stripped by the ram pressure 
or tidal forces of the host halo. Consequently, the 
star formation in the satellite can be suppressed or even quenched. 
Indeed, observations clearly show that satellite galaxies have
larger quenched fractions than centrals of the same stellar mass
\citep[e.g.,][]{Balogh00,vdBosch08,Wetzel12}. 

We use a simple $\tau$ model to describe the suppression of star formation 
in a satellite after it is accreted:
\begin{equation}\label{SFR_satellite}
 {\dot M}_{\star, st}(t) = {\dot M}_{\star}(t_{a}) 
                        \exp\left( -\frac{t-t_{a}}{\tau_{st}} \right) \, ,
\end{equation}
where $t_{a}$ is the time when the satellite is accreted into its
host, ${\dot M}_{\star}(t_{a})$ is the SFR of the
satellite galaxy at $t=t_a$, and $\tau_{st}$ is a time scale
characterising the decline of the star formation. We adopt the
following model for the characteristic time
\begin{equation}
 \tau_{st} = \tau_{st,0} \exp\left( -\frac{M_\star}{M_{\star,c}} \right) \,,
\end{equation}
where $\tau_{st,0}$ is the exponential decay time for a galaxy with a
stellar mass of $M_{\star,c}$, with both of these being free
parameters in our model.  The choice of $\tau_{st}$ is motivated by
the fact that massive galaxies tend to be more quenched in star
formation than low mass galaxies, independent of environment
\citep{Peng10, Wetzel13}. 
For central galaxies, this trend is naturally reproduced by assuming
that in massive halos the star formation efficiency decreases with
halo mass  (as long as $\alpha<0$).  For satellites, however,
such a halo-mass dependence of star formation efficiency does not
work, because the host subhalo mass is expected to decrease with time
owing to stripping. 

\subsection{Merging and stripping of satellite galaxies}

A halo with mass $M_{\rm sat}$ accreted by a larger (primary)
halo with mass $M_{\rm pry}$ will gradually sink towards the centre of
the primary halo owing to dynamical friction.  Using N-body
simulations, \citet{Boylan-Kolchin08} found that the dynamical
friction time scale, $\tau_{\rm merger}$, depends on the mass ratio
between the satellite halo and the host halo, as well as on the
orbital parameters of the satellite halo $\eta$ at the time of
accretion:
\begin{equation}\label{tau_merger}
  \tau_{\rm merger} 
  = 0.216 { \left(M_{\rm pry}/M_{\rm snd} \right)^{1.3} 
      \over \ln(1+M_{\rm pry}/M_{\rm snd}) } 
    \exp \left( 1.9 \eta \right) \tau_{\rm dyn}\,, 
\end{equation}
where $\tau_{\rm dyn}= r_{\rm vir}/V_{\rm vir}$ is the dynamical time
of the halo, with $r_{\rm vir}$ and $V_{\rm vir}$ being the virial
radius and virial velocity of the halo, respectively; $\eta$ is 
the orbital circularity, which is the ratio between the orbital
angular momentum and the orbital angular momentum of a circular
orbit with the same energy.  Following \citet{Zentner05}, we assume
$\eta$ to have the following distribution,
\begin{equation}
P(\eta)\propto \eta^{1.2}(1-\eta)^{1.2}\,.
\end{equation}
For each satellite galaxy at the time of accretion, we draw a value of
$\eta$ from this distribution and use it in Equation~(\ref{tau_merger}) 
to estimate a dynamical friction time scale.

We assume that a satellite galaxy merges with the central galaxy of
the primary halo in a time $\tau_{\rm merger}$ after accretion.  We
further assume that only a fraction $f_{\rm TS}$ of the stellar mass
of the satellite is added onto the central galaxy and the
rest of its stellar mass becomes halo stars. In our model we treat
$f_{\rm TS}$ as a free parameter.

\subsection{Spectral synthesis and metal enrichment}
\label{ssec:ssme}

Given a star formation model, we use the procedure described above,
together with a halo merger tree, to predict the star formation
history of a given galaxy. To convert that star formation history to a
stellar mass, which takes into account mass loss owing to stellar
evolution, and to calculate luminosities in different bands, we adopt
a Chabrier (2003) IMF and the stellar population synthesis model of
\citet{Bruzual03}. We correct the SMFs that we use as observational 
constraints to this IMF if they were originally estimated using 
a different IMF.

\begin{figure}
 \centering
 \includegraphics[width=0.9\linewidth, angle=270]{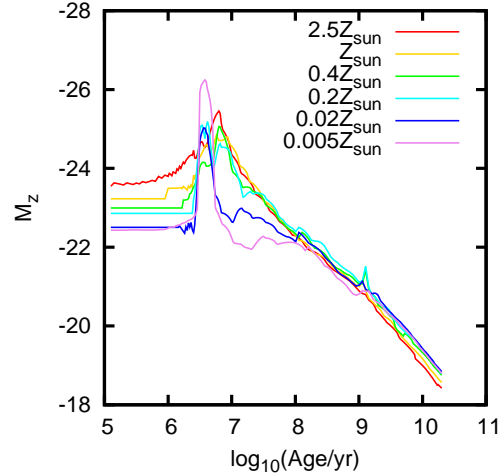}
 \caption{The $z$-band magnitude of a simple stellar population 
   as a function of age for different metallicities.}
\label{fig:zband}
\end{figure}

Since our model does not include the gas component in galaxies, we
cannot trace the chemical evolution of stars directly. Instead, we use
the metallicity - stellar mass relation observed for local galaxies
at all redshifts.  We adopt the mean relation based on the data of
\citet{Gallazzi05}, which can roughly be described as
\begin{equation}
 \log_{10} Z = \log_{10} Z_{\odot} + \frac{1}{\pi}
\tan \left[\frac{\log_{10}(M_{\star}/10^{10}M_{\odot})}{0.4}\right] - 0.3 \,.
\end{equation}
This observational relation extends down to a stellar mass of
$10^9M_{\odot}$ and has a scatter of $0.2\,{\rm dex}$ at the massive end
and of $0.5\,{\rm dex}$ at the low mass end. Fortunately, the $z$-band
luminosity, which we use to compare with observations, depends only
weakly on metallicity as shown in Figure~\ref{fig:zband}.  Hence, our
results are expected to be insensitive to the exact chemical evolution
model that we adopt.

\section{Observational Constraints and Likelihood Functions}
In this paper, we use `field galaxies' to refer to all galaxies 
independent of their environment. The data used in this paper 
are listed in Table~\ref{constraints}, including the SMF of field 
galaxies at 4 different redshift bins and the z-band CGLF.  

\begin{table*}
 \centering
 \caption{Observational Constraints. Column 2 lists the redshift ranges
 of the SMFs and the mean redshift of the z-band CGLF. 
 Column 3 lists the error model we use in the likelihood
 function: linear means normal distribution and log means log-normal distribution.
 Column 4 lists the sources of the data.}
 \begin{minipage}[l]{1.0\textwidth}
  \centering
  \begin{tabular}{llll}
 \hline
 \hline
         & Redshift  & Error model & Reference \\ 
 \hline
    SMF  & [0, 0.06] & linear      &\citet{Baldry12}   
  \vspace{3ex} \\ 
    SMF  & [1, 1.3]  & log         &\citet{PG08}
  \vspace{3ex} \\
    SMF  & [2, 3]    & log         &\citet{Marchesini09}
  \vspace{3ex} \\
    SMF  & [3.19, 4.73] & linear   &\citet{Stark09}
  \vspace{3ex} \\ 
    CGLF ($z$-band) & 0.1 (mean) &log  & \citet{Popesso06} \\    
 \hline
 \hline 
\end{tabular}
\end{minipage}
\label{constraints}
\end{table*}

\subsection{The stellar mass functions of field galaxies}

The SMF of the local Universe ($z\approx 0$) is from \citet{Baldry12}. 
The galaxies are selected from the SDSS DR6
down to a magnitude limit of $m_{r}\approx 19.8$ with a sky coverage
of $143\, {\rm deg}^{2}$. Redshift measurements for galaxies fainter
than the SDSS redshift survey limit are made with the Galaxy And Mass
Assembly (GAMA) survey.  Stellar masses for individual galaxies are
estimated from their $ugriz$ photometry using a spectral synthesis
model with the assumption of a \citet{Chabrier03} IMF.  Owing to
surface brightness incompleteness, the measured number density for
galaxies with masses below $10^8\Msunhh$ can only be considered as lower
limits. In our analysis we, therefore, only use data points above
$10^8\Msunhh$. All models presented here predict mass functions
that lie above the data points of \citet{Baldry12} for $M_\star
< 10^8 \Msunhh$, as required.

In addition to local galaxies, we also use the SMFs
of galaxies in the following three redshift bins: $1.0<z<1.3$ from
\citet{PG08}; $2.0<z<3.0$ from \citet{Marchesini09}; and $3.19<z<4.73$
from \citet{Stark09}.  Stellar masses in \citet{PG08} and
\citet{Stark09} were estimated using broad-band photometry assuming a
Salpeter IMF. Following \citet{Stark07}, we convert these masses into
those corresponding to a Chabrier IMF by dividing them by a factor of
$1.4$. The stellar masses in \citet{Marchesini09} were derived using a
pseudo-Kroupa (2001) IMF, but it was shown that adopting a Chabrier
(2003) IMF does not have a significant effect on their stellar mass
functions so we use their SMF directly without any
corrections.

To make predictions for the field galaxy SMF, we use 
5,000 dark matter halos drawn from a power law distribution
$f'(M_{\rm vir}) \propto \left[M_{\rm vir}/(\Msunh)\right]^{-1.5}$ in
the mass range $5\times10^{9}h^{-1} {\rm M}_{\odot}<M_{\rm vir} < 5
\times 10^{14} h^{-1} {\rm M}_{\odot}$. Including halos outside this
mass range does not change our results significantly. 
The fact that our merger trees do not resolve progenitor halos with
$M_{\rm vir}(z) < 2 \times 10^9 \Msunh$, implies that we implicitly
assume that star formation is suppressed in haloes with masses below
this `threshold'. In other words, our star formation model
(Eq.~[\ref{eqn_general02}]) should be augmented with $\dot{M}_\star
= 0$ for $M_{\rm vir}(z) < 2 \times 10^9 \Msunh$. When we calculate
the stellar mass/luminosity function of galaxies, the predicted number
of galaxies associated with a halo of mass $M_{\rm vir}$ is assigned a
weight $\omega = f_{\rm SMT}(M_{\rm vir})/f'(M_{\rm vir})$ where
$f_{\rm SMT}(M_{\rm vir})$ is the halo mass function from
\citet{Sheth01} for our adopted cosmology.

\subsection{The composite luminosity function of cluster galaxies}

As an additional constraint, we
also use the $z$-band composite luminosity function of cluster
galaxies obtained by \citet{Popesso06}.  The $z$-band luminosities
are less affected by dust extinction than their bluer counterparts,
making the comparison between our model and the data less affected by
dust corrections.  Furthermore, for a stellar population older than
100 Myr the predicted $z$-band luminosity depends only weakly on
metallicity, making our results less sensitive to the chemical
evolution model described above (see Fig.\,\ref{fig:zband}).  
Finally,the use of the composite luminosity function, a weighted average over
a number of individual clusters, reduces statistical uncertainties
arising from variances in the formation history of clusters.

The composite luminosity function of \citet{Popesso06} is based on 69
clusters selected from ROSAT X-ray data using cross identifications
with galaxies in the SDSS.  Once the luminosity functions of
individual clusters are known the composite luminosity
function can be evaluated using
\begin{equation}
N_{\rm cj} = {N_{\rm c0} \over m_{\rm j}} 
\sum_{i} {N_{\rm ij} \over N_{\rm i0}}\,,
\end{equation}
where $N_{\rm cj}$ is the number of galaxies in the $j$th bin of the
composite luminosity function, $N_{\rm ij}$ is the number of galaxies
in the $j$th bin contributed by the $i$th cluster, $N_{\rm i0}$ is the
normalisation factor for the $i$th cluster, which is the total number
of the member galaxies brighter than the magnitude limit $m_{\rm lim}$
at the cluster redshift, $m_{\rm j}$ is the number of clusters
contributing to the $j$th bin and
\begin{equation}
N_{\rm c0} = \sum_{i} N_{\rm i0}\,.
\end{equation} 
We follow the same method to calculate the composite luminosity
functions in our model predictions.

To use the CGLF as a constraint it is necessary to know the masses 
of those clusters accurately, since a systematic error in cluster mass 
can lead to an error in the amplitude of the predicted CGLF.  
Unfortunately, the mass estimates for the clusters in the \citet{Popesso06} 
sample are uncertain, and so it is dangerous to use the overall 
amplitude of the CGLF to constrain the model.  To bypass this problem,
we treat the ratio between the real mass of a cluster $M_{\rm real}$
and the measured value $M_{\rm obs}$,
\begin{equation}
  \label{eqn_err}
  e_{\rm M} \equiv \frac{M_{\rm real}}{M_{\rm obs}} \,, 
\end{equation}
as a free parameter, and we renormalise the luminosity function of an
individual cluster by $e_{\rm M}$ (i.e., we marginalise over
potential systematic biases in the inferred cluster masses).

To make predictions for the CGLF, we
generate $23$ merger trees for halos with a present-day mass
distribution similar to that observed. The mass resolution adopted 
for these merger trees is also $2 \times 10^9 \Msunh$. We confirmed
that this number of halos is sufficiently large so that the variance
between the different realisations is smaller than the uncertainties
in the observational data.

\subsection{The Likelihood Function and Sampling Algorithm}

The Likelihood function describes the probability of the data given
the model and its parameters.  A rigorous likelihood function includes
all the processes in the data acquisition. For the stellar mass
functions we study in this paper, the data acquisition process
includes deriving a stellar mass from multiband photometry using a
stellar population synthesis model, correcting for incompleteness,
weighting each galaxy sample according to its corresponding survey
volume, and so on.  As a result, the uncertainties in individual
stellar mass bins are not independent.  As is demonstrated in
\citet{Lu11}, the covariance for binned data may change the posterior
distribution substantially.  Unfortunately, the covariance matrix in
the data used here is not available and we have to assume that the
stellar mass function in different bins is independent and that the
likelihood is approximated by a Gaussian function. 
As shown in Appendix~\ref{sec_like}, 
with these assumptions the likelihood function can be written as 
\begin{equation}
\label{like}
\ln L(\Phi_{\rm obs}|\theta)
= C - {1 \over 2} \sum_{i} 
{\left[\Phi_{i,{\rm obs}} - \Phi_{i,{\rm mod}}(\theta) \right]^{2} 
 \over \sigma_{i,{\rm obs}}^{2} },
\end{equation}
where $\Phi_i$ and $\sigma_i$ are either defined in linear space or
logarithmic space, depending on the observational data (see
Table~\ref{constraints}), and $C$ is an unimportant factor.

To efficiently sample the high dimensional parameter space,
we make use of the MULTINEST method developed by \citet{Feroz09},
which implements the nested sampling algorithm of \citet{Skilling06}.
We have compared this method with the MCMC 
implemented in \citet{Lu11}, Tempered Differential Evolution.
Both methods are designed to deal with probability 
distributions with multiple modes and strong degeneracies between 
model parameters. For the problem in this paper, we found the two 
give identical results, but that the number of likelihood evaluations 
required by MULTINEST is smaller by more than a factor of 10.
A more detailed description of the method is given in 
Appendix~\ref{sec_multinest}. 

\section{Models of star formation in dark matter halos}

\begin{table}
\caption{Summary of Posterior Simulations. Column 3 is the natural log of 
the marginalised likelihood given the models (Column 1) and the data (Column 2).
The ratio between the marginalised likelihood is the Bayes Factor, 
which is the odds that the given data favour one model over the other.}
\begin{tabular}{lll}
\hline
\hline
 Model        & Constraints   & Marginalised Likelihood \\
              &               & (natural log)           \\
\hline
 Model I      & SMF($z=0$)    & $-22.2$
 \vspace{3ex} \\
 Model I      & SMF           & $-120 $ 
 \vspace{3ex} \\
 Model II     & SMF($z=0$)    & $-21.0$
 \vspace{3ex} \\
 Model II     & SMF           & $-31.2$
 \vspace{3ex} \\
 Model II     & SMF + CGLF    & $-89.7$
 \vspace{3ex} \\
 Model IIb    & SMF           & $-55.5$
 \vspace{3ex} \\ 
Model III     & SMF           & $-30.6$
 \vspace{3ex} \\ 
 Model III    & SMF + CGLF    & $-63.3$
 \vspace{3ex} \\
 Model IV     & SMF + CGLF    & $-58.4$\\   
\hline
\hline 
\end{tabular}
\label{summary}
\end{table}

We explore a series of nested model families, based on the generic
parametrisation given by Equation~(\ref{eqn_general02}).  At each step
we increase the complexity of the model and check whether the fit to
the observational constraints is acceptable and whether any
improvement to the fit is sufficient to justify the increased
complexity by comparing the posterior predictions with the
constraints.  In addition, we also make use of the Bayes Factor to
avoid developing an overcomplicated model (Table~\ref{summary}).  In
Bayesian statistics, the Bayes Factor is
\begin{equation}
\label{bayesfactor}
 K = \frac{p(M_{\rm a}|D)}{p(M_{\rm b}|D)}\, ,
\end{equation}
where $p(M|D)$ is the probability of model $M$ given data $D$. 
It is obtained by marginalising over all the model parameters 
of the posterior distribution. The advantage
of the Bayes factor is that it automatically includes a penalty for too
much complexity in the model.
The values of $\ln\left[p(M|D)\right]$ for all the models discussed
 in the text are listed Table~\ref{summary}.  
 
The model families explored are summarised in Table~\ref{model}, and
the prior ranges of the model parameters are also given in the table. For
some parameters, the prior ranges are motivated by other studies. For
instance, $\alpha$ is only allowed to be negative because of the
strong quenching of star formation found in massive galaxies.  Some
prior choices are made to avoid unphysical modes.  
A prior range that is too wide 
sometimes contains unphysical modes with high probability,
making the physical modes negligible. Such priors have to be excluded.
In general, the prior covers a pretty large range of possibilities.
The median and 95\% interval of the posterior model parameters are
also given in the table.

\begin{table*}
\caption{A list of the model parameters of the four main model families
         considered. The prior ranges, the medians and the 95\% credible intervals
         of the model parameters are listed. $M_{\rm c}$ is in units 
         of $10^{10}\Msunh$, and $M_{\rm *,c}$ is in units of $10^{10}\Msunhh$.
        }
\setlength\tabcolsep{2pt}
\begin{tabular}{cccccccc}
   \hline \hline
   \multicolumn{2}{c}{\textbf{Model I, SMF($z=0$)}}   & 
   \multicolumn{2}{c}{\textbf{Model II, SMF}}  & 
   \multicolumn{2}{c}{\textbf{Model III, SMF+CGLF}} & 
   \multicolumn{2}{c}{\textbf{Model IV, SMF+CGLF}}  \\
   \hline
   
   Parameter   &median       & Parameter   & median       & Parameter   & median       & Parameter   & median    \\
   prior       &$95\%$ range & prior       & $95\%$ range & prior       & $95\%$ range & prior       & $95\%$ range    \\
   \hline

   $\alpha$ & $-1.6$         & $\alpha_{0}$ & $-3.6$         & $\alpha_{0}$ & $-3.1$         & $\alpha_{0}$  & $-4.2$         \\ 
   $[-5,0]$ & $[-2.5, -1.1]$ &  $[-5,0] $   & $[-4.9, -2.1]$ & $[-5,0]$     & $[-4.9, -1.6]$ & $[-5,0]$      & $[-4.9,-2.9]$  \\  [2ex]
 
            &                & $\alpha'$  & $-0.72$         & $\alpha'$  & $-0.69$        & $\alpha'$ & $-1.2$         \\
            &                & $[-2,0]$   & $[-1.09,-0.54]$ & $[-2,0]$   & $[-0.89,-0.49]$& $[-2,0]$  & $[-1.5,-0.88]$ \\ [2ex]

   $\beta$  & $3.5$          & $\beta$    & $1.8$          & $\beta$   & $1.8$         & $\beta$  & $2.9$          \\
   $[0, 5]$ & $[1.3, 4.9]$   & $[0,5]$    & $[0.08, 3.5]$  & $[0,5]$   & $[1.1, 3.7]$  & $[0,5]$  & $[1.8,4.6]$    \\ [2ex]

   $\gamma$ & $0.92$        & $\gamma$    & $1.9$          & $\gamma_{a}$ & $2.6$         & $\gamma_{a}$   & $2.5$         \\
   $[-1,3]$ & $[0.05,2.5]$  & $[-1,3]$    & $[0.24, 2.8]$  & $[-1,3]$     & $[1.5, 2.9]$ & $[-1,3]$       & $[0.95,3.0]$ \\ [2ex]

            &                &          &                &              &                & $\gamma''$     & $-0.55$       \\
            &                &          &                &              &                & $[-1,1]$       & $[-0.98,0.48]$ \\ [2ex]

            &                &          &              & $\gamma_{b}$  & $-0.88$        & $\gamma_{b}$ & $-0.84 $              \\
            &                &          &              & $[-1,1]$      & $[-0.99, -0.41]$ &  $[-1,1]$    & $[-0.99, -0.42]$    \\ [2ex]

            &                &          &                & $\gamma'$ & $-4.3$         &  $\gamma'$  & $-4.4$         \\
            &                &          &                & $[-5,0]$  & $[-4.9, -2.4]$ &  $[-5,0]$   & $[-4.9,-2.6]$  \\ [2ex]

            &                &          &                & $z_{\rm c}$  &  $2.1$         & $z_{\rm c}$    & $1.8$        \\
            &                &          &                & $[0,10]$     & $[1.5, 2.7]$   & $[0,10]$       & $[0.44, 2.4]$ \\ [2ex]  

  $\log_{10}(M_{\rm c})$ & $1.4$        & $\log_{10}(M_{\rm c})$ & $1.9$        & $\log_{10}(M_{\rm c})$ & $1.8$        & $\log_{10}(M_{\rm c,0})$ & $1.6$        \\
  $[0,4]$          & $[1.2, 1.8]$ & $[0,4]$          & $[1.5, 2.2]$ & $[0,4]$          & $[1.4, 2.1]$ & $[0,4]$            & $[1.4,1.8]$ \\[2ex]

                   &             &                 &             &               &             & $\mu$ & $-0.09$         \\
                   &             &                 &             &               &             &$[-1,1]$& $[-0.91,0.94]$ \\[2ex]

 $\log_{10}({\cal R})$ & $-0.83$        & $\log_{10}({\cal R})$ & $-0.96$          & $\log_{10}({\cal R})$ & $-1.1$         & $\log_{10}({\cal R}_{\rm 0})$ & $-0.91$  \\
 $[-2,0]$        & $[-1.8,-0.29]$ & $[-2,0]$        & $[-1.9, -0.20]$  & $[-2,0]$        & $[-1.5,-0.56]$ & $[-2,0]$        & $[-1.4,-0.17]$ \\[2ex]

                 &             &               &               &                &             & $\rho$ & $0.18$           \\
                 &             &               &               &                &             &$[-1,1]$&$[-0.62,91]$ \\ [2ex]

 $\log_{10}({\cal E})$ & $-0.21$        & $\log_{10}({\cal E})$ & $-0.27$        & $\log_{10}({\cal E})$ & $-0.32$         & $\log_{10}({\cal E})$ & $0.36$        \\
 $[-2,1]$        & $[-0.52,0.21]$ & $[-2,1]$        & $[-0.74,0.09]$ & $[-2,1]$        & $[-0.55,0.03]$ & $[-2,1]$        & $[0.04,0.74]$\\[2ex]

                 &             &           &             &           &             &   $\kappa'$    & $-1.3$       \\
                 &             &           &             &           &             &  $[-1.5,1.5]$  & $[-1.5,-0.95]$ \\[2ex]

   $\log_{10}(H_{0}\tau_{\rm st,0})$ & $-0.71$          & $\log_{10}(H_{0}\tau_{\rm st,0})$ & $-1.1$         & $\log_{10}(H_{0}\tau_{\rm st,0})$ & $-1.37$         & $\log_{10}(H_{0}\tau_{\rm st,0})$ & $-0.86$         \\
   $[-2,-0.8]$  & $[-1.9,-0.80]$  & $[-2,-0.8]$  & $[-1.94,-0.80]$ & $[-2,-0.8]$  & $[-1.9,-0.83]$  & $[-2,-0.8]$  & $[-1.1,-0.80]$ \\ [2ex]

   $\log_{10}(M_{\rm *,c})$ & $-1.1$         & $\log_{10}(M_{\rm *,c})$ & $-1.4 $         & $\log_{10}(M_{\rm *,c})$ & $-1.4$         & $\log_{10}(M_{\rm *,c})$ & $-1.5$         \\
   $[-2,1]$       & $[-2.0, 0.91]$ & $[-2,1]$       & $[-1.94, 0.90]$ & $[-2,1]$       & $[-1.9, 0.78]$ & $[-2,1]$       & $[-2.0,0.17]$ \\ [2ex]

  $f_{\rm TS}$ & $0.65$         & $f_{\rm TS}$ & $0.099$         & $f_{\rm TS}$ & $0.13$         & $f_{\rm TS}$ & $0.11$         \\
  $[0,1]$      & $[0.25, 0.95]$ & $[0,1]$      & $[0.004,0.28]$  & $[0,1]$      & $[0.03, 0.27]$ & $[0,1]$      & $[0.01, 0.35]$  \\ [2ex]

               &      &               &      & $\log_{10}(e_{\rm M})$  & $0.22$         & $\log_{10}(e_{\rm M})$  & $0.17$         \\
               &      &               &      & $[-0.5,0.5]$ & $[0.14, 0.27]$  & $[-0.5,0.5]$ & $[0.11, 0.23]$ \\
  \hline
  \hline
\end{tabular}
\label{model}
\end{table*}

\subsection{The Fiducial Model Family (Model I)}

\begin{figure*}
   \centering
   \includegraphics[width=0.8\linewidth,angle=270]{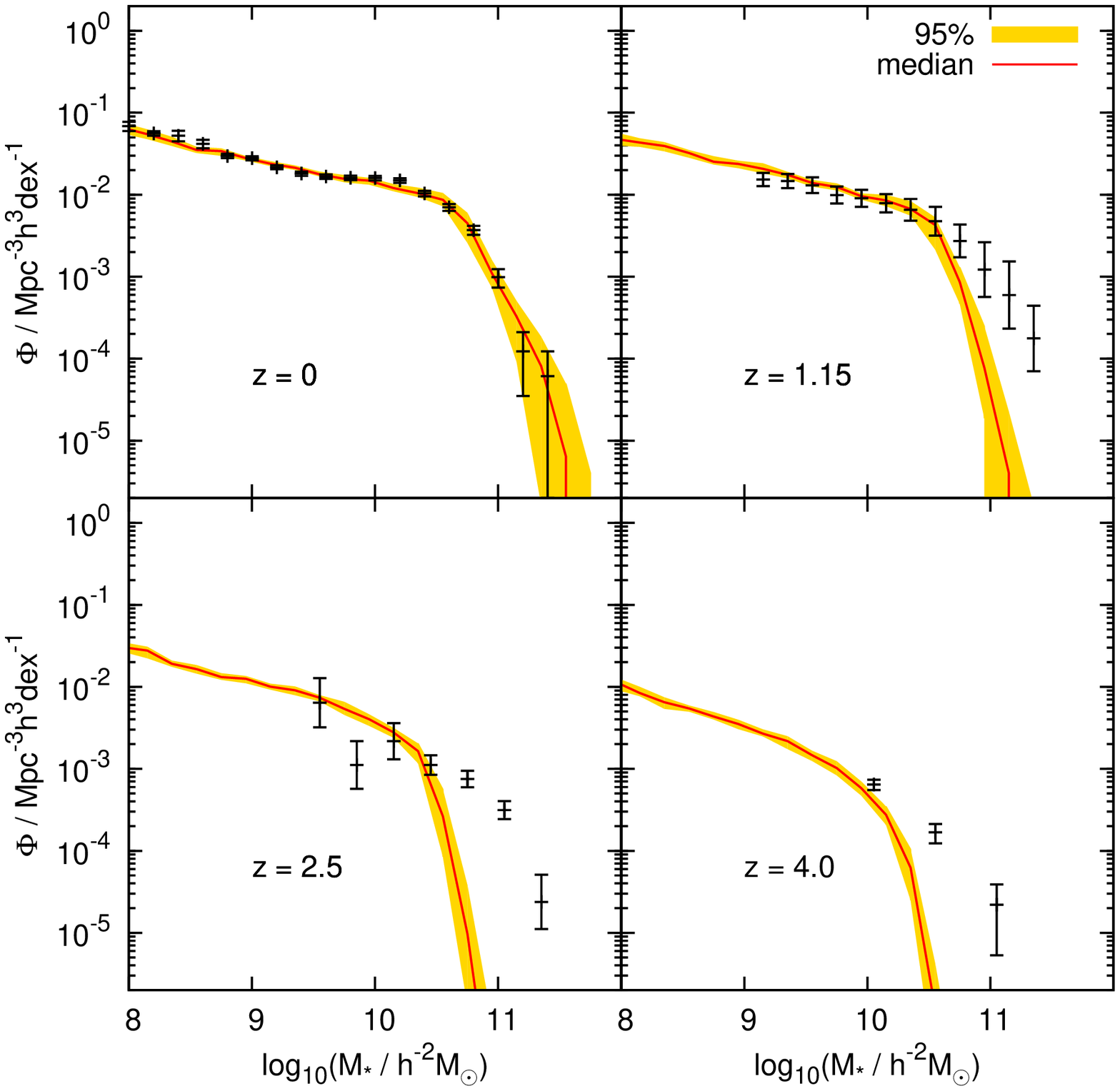}
   \caption{The posterior predicted SMF of Model I
     constrained by the SMF at $z\approx0$.  The yellow
     bands encompass the 95\% credible intervals of the posterior
     distribution, while the red solid lines are the medians. The black
     data points with error bars are the observational data. Note that
     the observational data at $z=1.15$, $2.5$ and $4.0$ are {\it not}
     used as constraints.}
   \label{modelI_a_smf}
\end{figure*}

\begin{figure*}
   \centering
   \includegraphics[width=0.8\linewidth,angle=270]{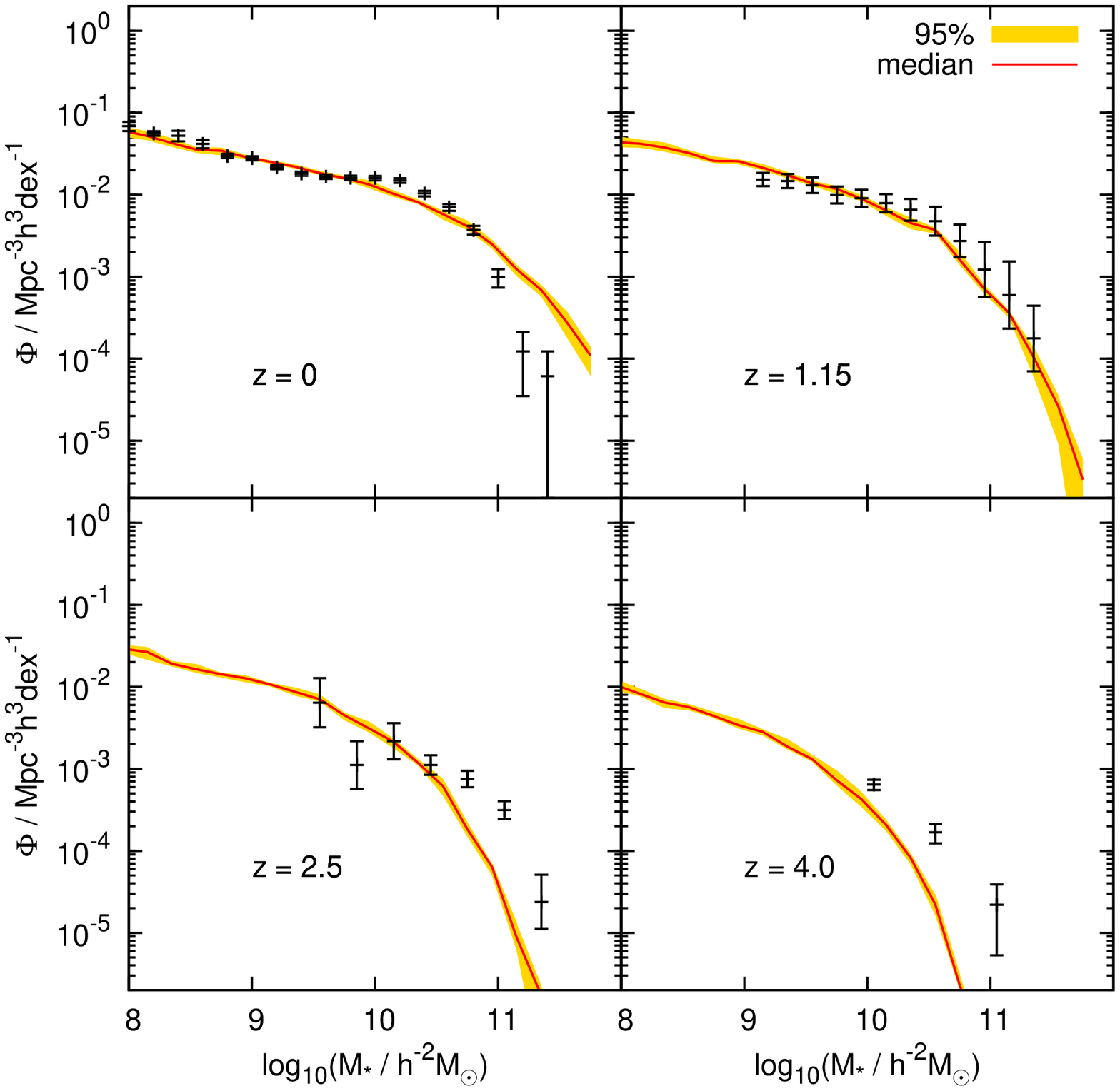}
   \caption{The predicted SMF of Model I constrained
     with SMFs at different redshifts.  The yellow
     bands encompass the 95\% credible intervals of the posterior
     distribution, and the red solid lines are the medians. The black
     data points with error bars are the observational constraints.}
   \label{modelI_b_smf}
\end{figure*}

We start with the simplest case in which all the parameters in
Equation~(\ref{eqn_general02}) are assumed to be time-independent, so
that the SFR is determined completely by the halo mass
and by the dynamical time scale of the halo at the time in question.
We call this Model I.  First, we use only the SMF at $z\approx0$
\citet{Baldry12} to constrain the model parameters.  
Figure~\ref{modelI_a_smf} shows that Model I fits the
constraining data well (see the first panel).  For reference, the
constrained model parameters are listed in Table~\ref{model}.  We
accomplish this fit using only nine parameters compared to the
thirteen (more physically motivated) parameters of the SAM used to fit
similar data in \citet{Lu11}. However, as one can see from the other
three panels of Figure~\ref{modelI_a_smf}, the posterior of the model
predicts too few massive galaxies at high $z$ compared to the
observational data.  Next, we use the SMFs at all
four different redshifts, as listed in Table~\ref{constraints}, to
constrain the model parameters.  The resulting fits to the data
are shown in Figure~\ref{modelI_b_smf}.  This model family still
fails to reproduce the SMFs at the bright end: it
overestimates the number density of bright galaxies at $z=0$ but
underestimates the number density at $z=2.5$ and $z=4$. This suggests
that massive galaxies form too late, and that the SFR at high-$z$ 
in massive halos needs to be boosted.

\subsection{Fixing the Bright-End Problem (Model II)}
\label{ssec_ModelII}

As an attempt to fix the bright-end problem identified above, we
consider a second model family (Model II) that allows $\alpha$ to be
redshift-dependent. We assume that the redshift-dependence be given by
the following power law,
\begin{equation}\label{eqn_alpha}
  \alpha = \alpha_{0} (1+z)^{\alpha'} \,, 
\end{equation}
where both $\alpha_0$ and $\alpha'$ are introduced as new free
parameters. The SFR in massive halos with $M_{\rm vir}>M_{\rm c}$ 
will be enhanced at high-$z$ if $\alpha'$ is a negative number.

\begin{figure*}
  \centering
  \includegraphics[width=0.8\linewidth,angle=270]{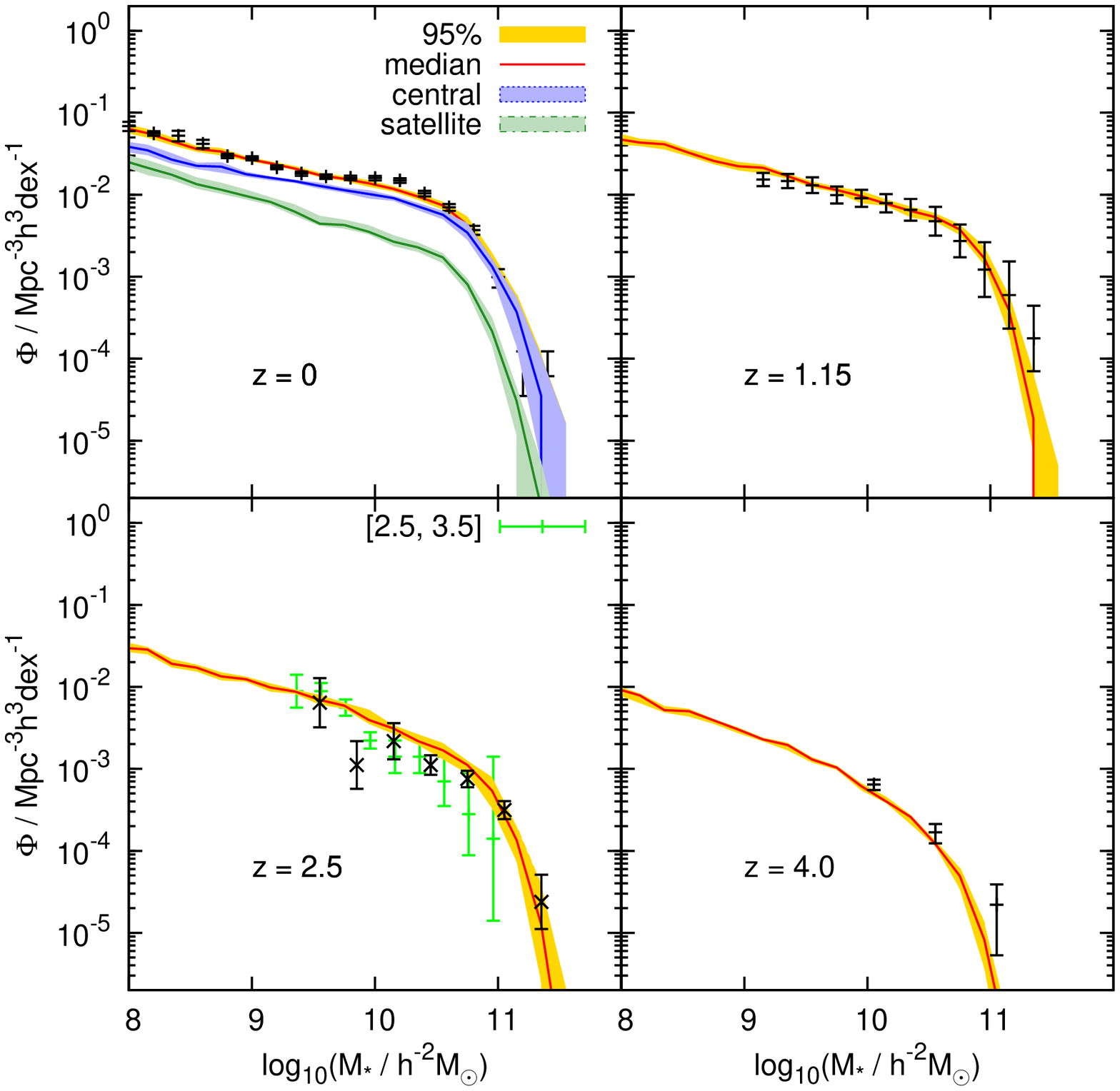}
  \caption{The predicted SMFs of Model II
    constrained with the SMFs listed in
    Table~\ref{constraints}.  The yellow bands encompass the 95\%
    credible intervals of the posterior distribution and the red solid
    lines are the medians. The black data points with error bars are
    the observational constraints. For $z\approx 0$ galaxies (upper
    left panel) , the contributions of satellites and centrals are
    plotted separately.}
\label{modelII_smf}
\end{figure*}

Once again we use the four SMFs listed in
Table~\ref{constraints} as observational constraints.
Figure~\ref{modelII_smf} compares the posterior predictions with the
constraining data.  This model family matches the observational data
over the entire redshift range from $z= 0$ to $z=4$.  The natural log
of the Bayes ratio between Model II over Model I (constrained by the
same data sets) is $-31.2-(-120) = 88.8$, 
making the odds of preferring Model II over Model I given the data 
a whopping $e^{88.8}$ to one (Table~\ref{summary}). 
At $z=0$ the contribution of satellites
to the total mass function is lower than that of the central
galaxies, with a satellite to central ratio of about 1/2 at the faint
end, decreasing to about 1/3 for higher stellar masses (see the top
left panel).  These results are in good agreement with those obtained
by Yang et al. (2008) based on galaxy groups selected from the SDSS
\citep[see also][]{Mandelbaum06,Cacciato13}. 

The credible intervals for the parameters are listed in Table~\ref{model}.
$\alpha^{'}$ lies between $-1.09$ and $-0.54$, which means
significant evolution in the SFR of massive
central galaxies. The constrained $f_{\rm TS}$ lies below $0.3$,
indicating that less than $30\%$ of the stars in disrupted 
satellites will be accreted by the centrals, with the rest 
of them remaining as halo stars (or intracluster stars in the case
of massive halos). This is consistent with early studies.
\citet{Yang12} find that in halos as massive as $10^{14}\Msunh$,
the total mass in disrupted satellites exceeds the mass 
of centrals, giving rise to intracluster stars. 
Also \citet{Purcell07}, \citet{Conroy07}, and \citet{Watson12} 
suggest that the majority of the stars in the satellites will become 
halo stars when the host subhalos are disrupted.

\begin{figure*}
\centering
\includegraphics[width=0.45\linewidth,angle=270]{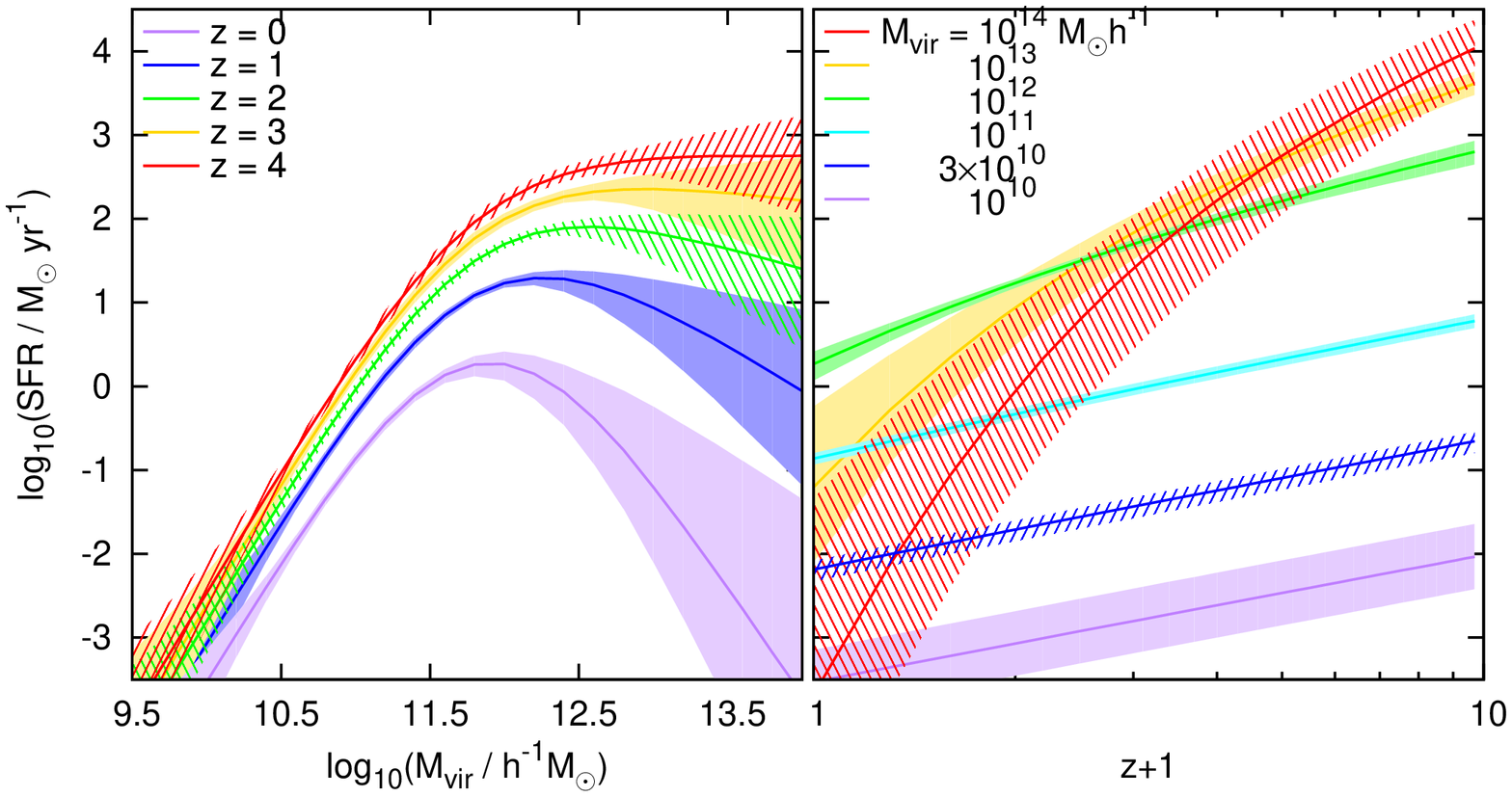}
\caption{The SFR as a function of halo mass (left panel) 
       and redshift (right panel) of Model II, constrained by the four SMFs.
       The solid lines are the medians of the posterior predictions
       and the bands are the 95\% credible intervals.}
\label{modelII_map}
\end{figure*}

Figure~\ref{modelII_map} shows the posterior prediction of Model II
for the SFR as a function of halo mass at different
redshifts (left panel) and as a function of redshift for different
halo mass bins (right panel).  The SFR peaks at around
$10^{12}\Msunh$ at each redshift. In halos with low masses the SFR increases
rapidly with halo mass, roughly as ${\rm SFR}\propto M_{\rm
  vir}^{2.5}$, quite independent of redshift. The SFR decreases with
halo mass for halos above $10^{12}\Msunh$.  The decrease is more
pronounced at low redshifts, becomes weaker as one goes to higher
redshifts, and is quite weak by $z=4$.  It should be noted, however,
that at $z\approx 4$ the SFR beyond $10^{13}\Msunh$ can only be
considered as an extrapolation because the number density of such
high-mass halos is extremely small at such high redshifts. For
halos with masses $\sim 10^{12}\Msunh$, the SFR increases with
redshift roughly as $(1+z)^{2.3}$, which is roughly proportional to 
the halo mass accretion rate as a function of redshift
\citep{Neistein08,Genel08,McBride09}. This increase with redshift is
faster for more massive halos, while for $M_{\rm vir}< 10^{11}\Msunh$,
${\rm SFR}\propto (1+z)^{1.5}$.

\begin{figure*}
  \centering
  \begin{minipage}[c]{0.45\linewidth}
     \includegraphics[width=\linewidth,angle=270]{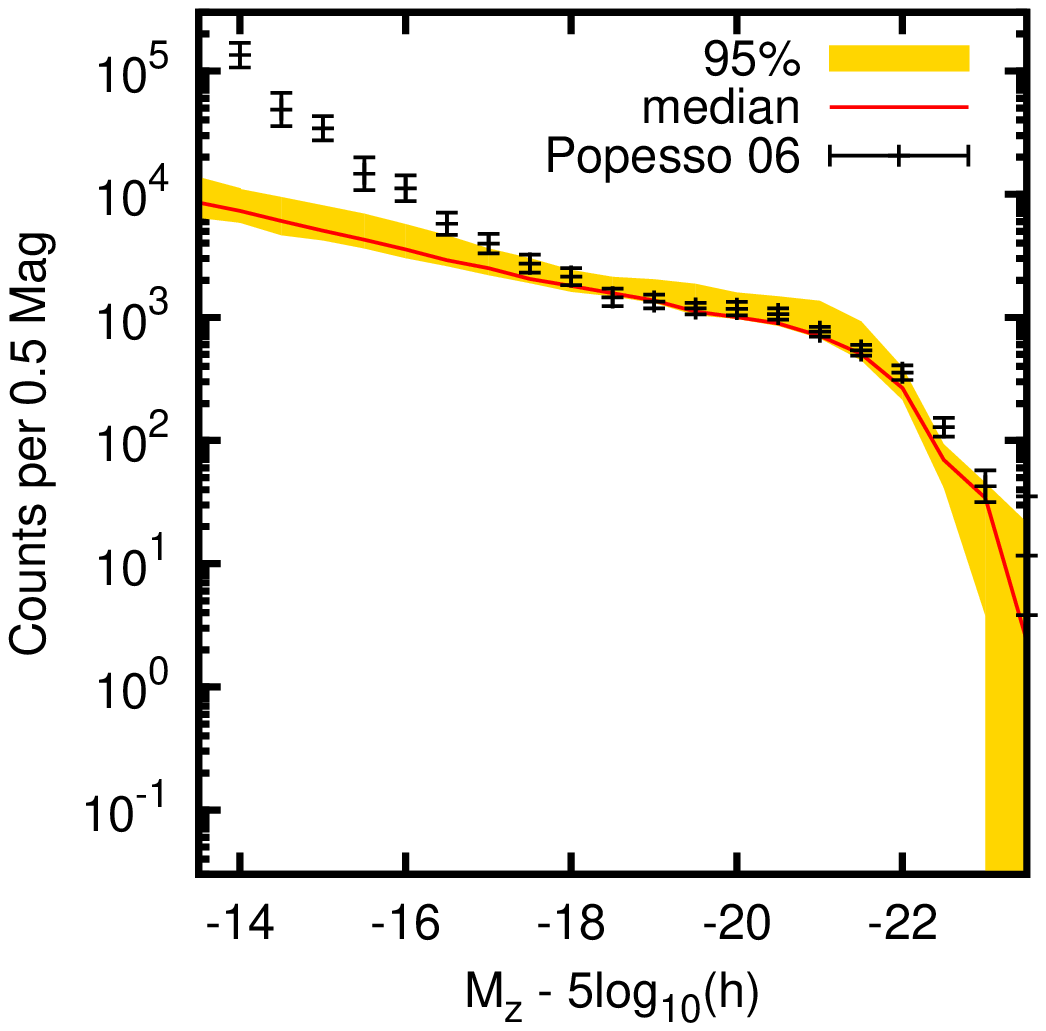}
  \end{minipage}
  \begin{minipage}[c]{0.45\linewidth}
     \includegraphics[width=\linewidth,angle=270]{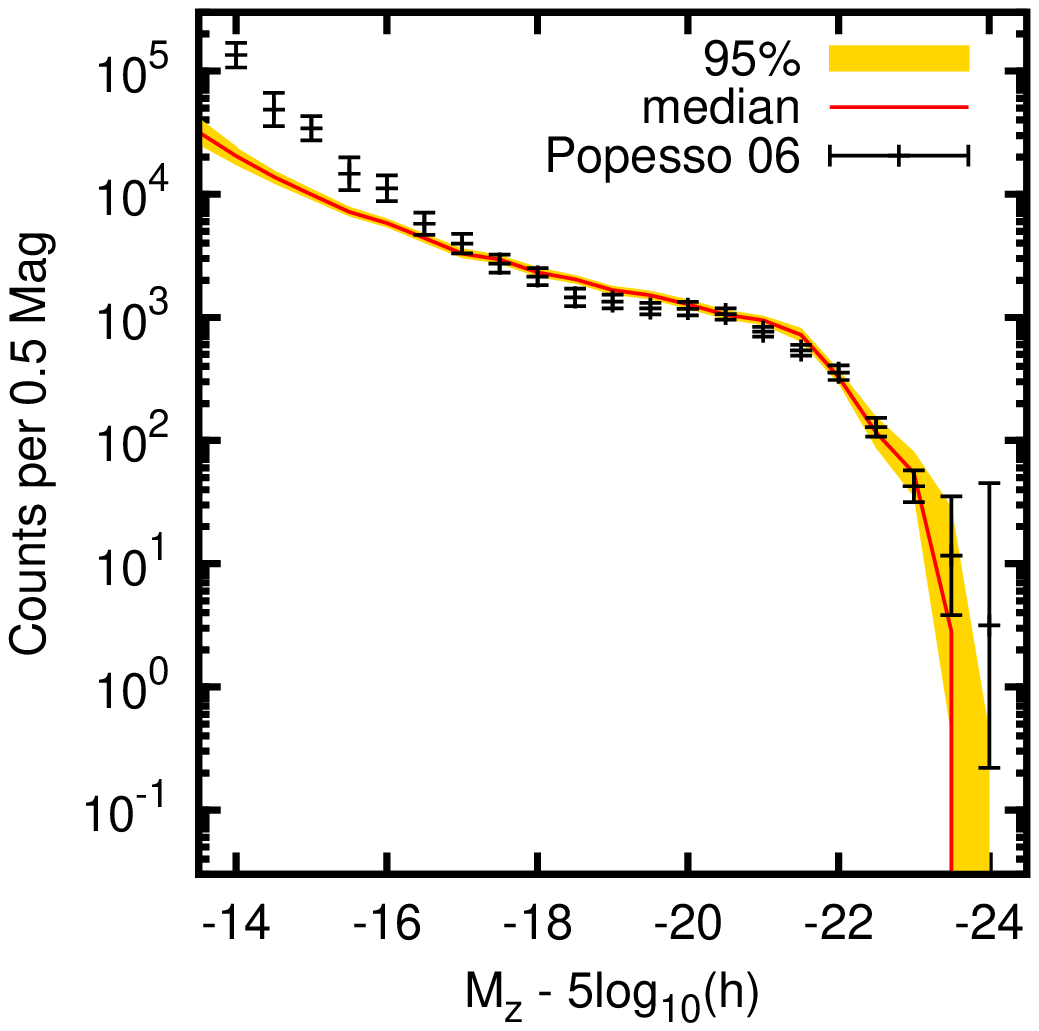}
  \end{minipage}
  \caption{Left panel: The posterior prediction of Model II constrained 
    by the SMFs compared with the CGLF of Popesso et al. (2006).
    The yellow band encompasses 95\% of the posterior distribution, 
    and the red line is the median. The observational data are shown 
    as data points with errorbars. 
    Right panel:  The same as the left panel but now Model II is constrained
    both by the SMFs and the observed $z$-band CGLF.}
  \label{modelII_clf}
\end{figure*}

In the following we examine how Model II matches the z-band CGLF.
The posterior prediction is shown in the left panel of Figure~\ref{modelII_clf}.
The normalisation of the prediction is adjusted to match
the observations at $M_{z}-5\log_{10}(h) \approx
-20$ to account for a potential systematic bias in the cluster masses. 
The model prediction is consistent with the observational data at the bright end but it
under-predicts the number of dwarf galaxies with $M_{z}-5\log_{10}(h) >
-17$. In particular, the predicted faint end of the luminosity
function is roughly a power law, in contrast with the observational
data that shows a significant upturn. To test whether the model family
Model II can accommodate the observed luminosity function of cluster
galaxies, we carry out a new inference, this time including the
$z$-band luminosity function as an additional data constraint.  The
predicted CGLF still fails to match the observed one at the faint end, 
as shown in the right panel of Figure~\ref{modelII_clf}. 
Thus, it is unlikely to find a model in the parameter
space of Model II to simultaneously match the SMFs and CGLF.

\subsection{
Fixing the faint end problem of the cluster galaxy
  luminosity function (Model III) }\label{model-III}

There are at least two possibilities that could lead to the
differences between the SMF of field galaxies and the CGLF.  
Since the over-abundance of dwarf
galaxies is more prominent in galaxy clusters, some environmental
effects specific to high-density environments may cause the strong
upturn.  However, many processes such as tidal stripping and tidal
disruption are destructive, which would suppress the number of
satellite galaxies rather than enhance it. It is possible that
relatively massive galaxies could have experienced significant mass
loss, moving them to lower masses. However, since more massive
galaxies are less abundant, it is difficult to make such a scenario
work in detail. Thus, unless environmental effects in clusters operate
in a way very different from what is generally believed, it is
difficult to explain the over-abundance of dwarf galaxies in clusters
with such effects.

Another possibility is that the environmental dependence of the
luminosity function may be the result of some time-dependent
processes.  Indeed, galaxies in a cluster are expected to form early
in their progenitor halos. Thus, if star formation in dark matter
halos were different at high redshift when the majority of cluster
galaxies formed, then the luminosity functions of cluster galaxies and
field galaxies could show different behaviours owing to their
systematically different formation times.  One concrete example is the
preheating model proposed in \citet{Mo02}, where the intergalactic
medium (IGM) is assumed to be preheated at some high redshift, so that the
star formation in dark matter halos proceeds differently before and
after the preheating epoch. Such preheating may owe to the formation
of pancakes, as envisaged in \citet{Mo05}, owe to an episode of
starbursts and AGN activity \citep{Mo02}, or owe to heating by
high-energy gamma rays generated by blasars, as envisaged in \citet{Chang11}. 
In all these cases, the preheating is expected to occur around
$z\approx 2$ and the preheated entropy of the IGM is a few times
$10\,{\rm KeV\,cm^2}$.  In what follows, we consider a generic model
family (Model III) inspired by the physical processes discussed above.
We allow $\gamma$, which controls star formation in low-mass halos, to
be time-dependent in such a way that it changes from $\gamma_{\rm b}$
at high-$z$ to $\gamma_{\rm a}$ at low-$z$, with a transition redshift
$z_{\rm c}$. Specifically we assume that
\begin{equation}\label{eqn_gamma}
  \gamma = 
  \begin{cases}
    \gamma_{\rm a}   \, & \text{if}\, z < z_{c} \\
    (\gamma_{\rm a}-\gamma_{\rm b})
    \left(\frac{z+1}{z_{c}+1}\right)^{\gamma'} + 
     \gamma_{\rm b}  \, & \text{otherwise}\,.
  \end{cases}
\end{equation}
Note that Model II is a special case of Model III, with $\gamma'=0$.
If $z_c=0$ and $\gamma_{\rm b}=0$ then
$\gamma$ is a simple power law of $(1+z)$ with an index of $\gamma'$.

\begin{figure*}
  \centering
  \includegraphics[width=0.8\linewidth,angle=270]{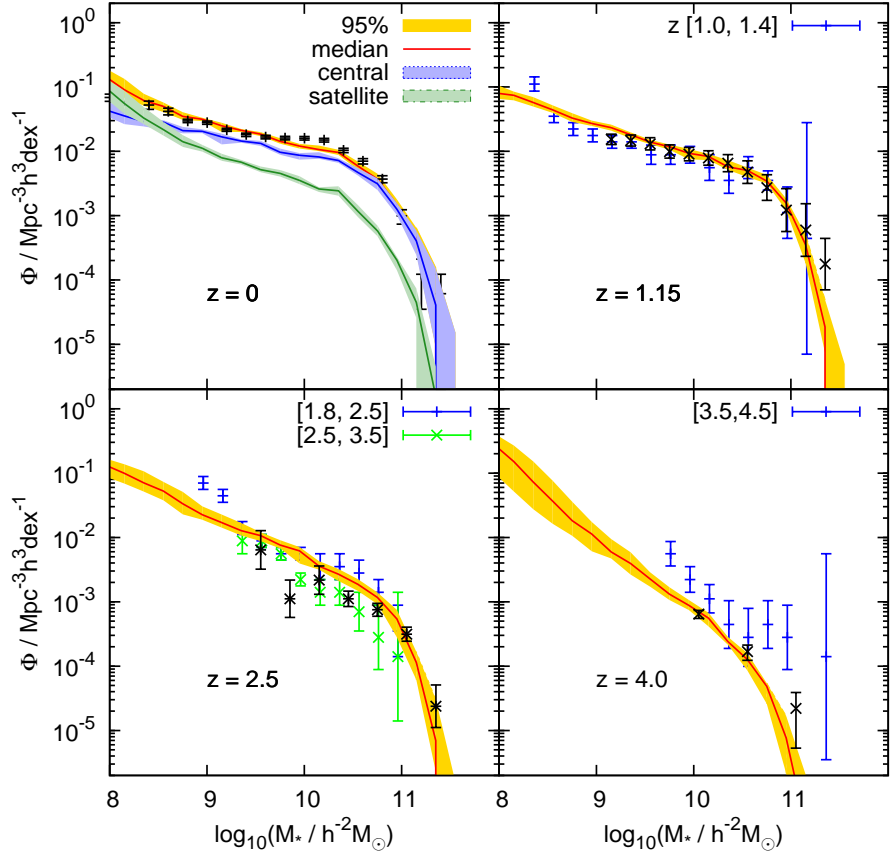}
  \caption{The SMFs predicted by the posterior of Model III 
    constrained by both the SMFs and the observed $z$-band CGLF.  
    The yellow bands encompass the 95\% credible
    intervals of the posterior distribution, and the red solid lines
    are the medians. The black data points with error bars are the
    observational constraints.  For comparison, we also plot, as
    coloured points, the results of \citet{Santini12} obtained from
    recent WFC3 data.  For $z\approx 0$ galaxies (upper left panel) ,
    the contributions of satellites and centrals are plotted
    separately.}
\label{modelIII_smf}
\end{figure*}

\begin{figure}
  \centering
  \includegraphics[width=0.9\linewidth,angle=270]{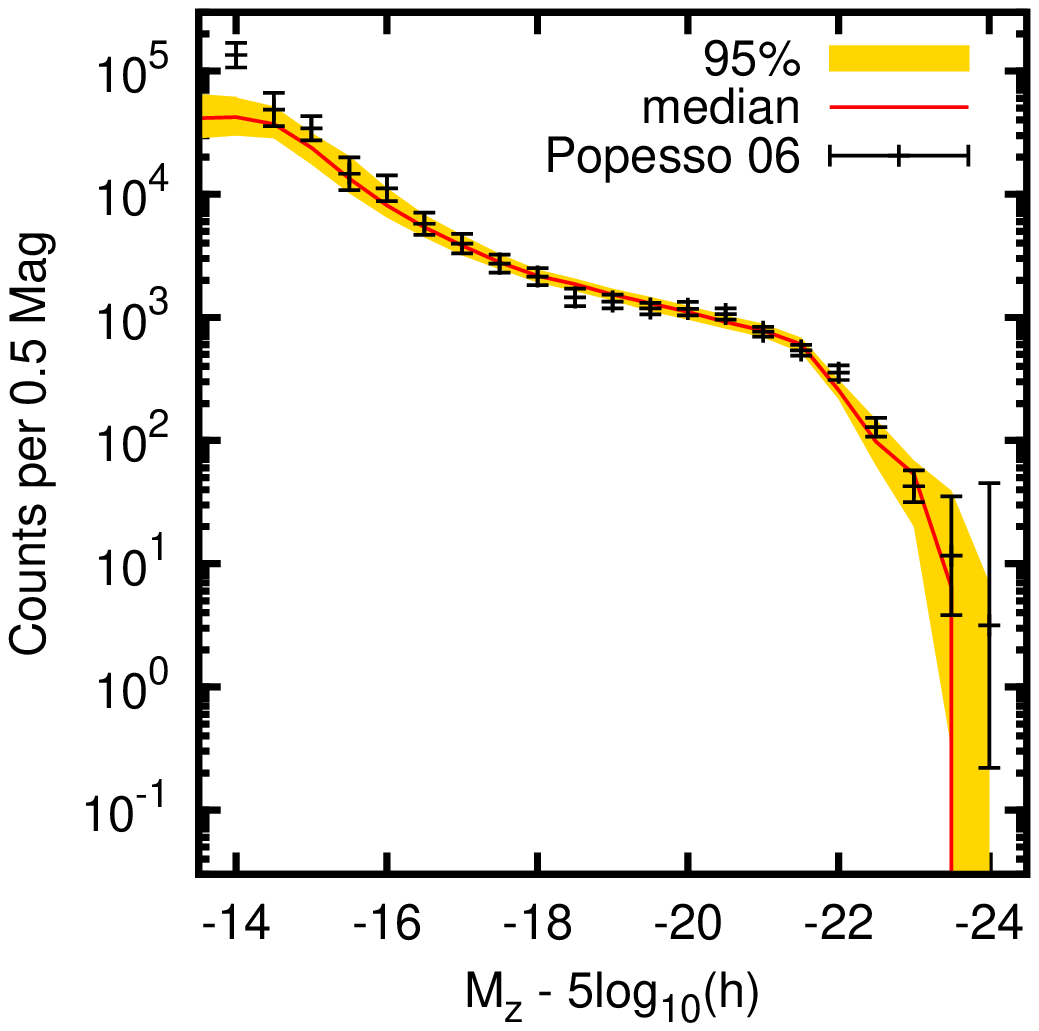}
  \caption{The CGLF predicted by the posterior of Model III 
    constrained by both the SMFs and the observed $z$-band CGLF.  
    The yellow band encompasses the 95\% credible
    interval of the posterior distribution, and the red solid line is
    the median. The black data points with error bars are the
    observational constraints.}
\label{modelIII_lf}
\end{figure}

Before we add this to Model II, thereby creating Model III, we want to
explore the possibility that adding this behaviour to Model I and
not allowing the massive end slope $\alpha$ to depend on redshift might
also provide a viable fit to the SMFs at the four
redshifts.  We refer to this as Model IIb. We perform an inference with
Model IIb using just the SMFs as data constraints.
As one can see from Table~\ref{summary}, the SMFs prefer 
Model II over Model IIb by a probability of $e^{24.3}$ to one, but
they still prefer Model IIb over Model I.

Figures~\ref{modelIII_smf} and \ref{modelIII_lf} compare the
posterior predictions with the constraining data, which are the  
SMFs and the CGLF. 
We see that Model III can accommodate both observational data sets.  
In particular, the upturn in the faint end of the CGLF
is well reproduced (see Fig.~\ref{modelIII_lf}).
The two data sets prefer Model III over Model II 
by a probability of $e^{26.4}$ to one (Table~\ref{summary}).
Comparing Figure~\ref{modelIII_smf} with Figure~\ref{modelII_smf} 
shows that the SMFs predicted by Model III are steeper than those
predicted by Model II, particularly at high redshift. 
The more recent results obtained by \citet{Santini12}
from recent WFC3 data, which are also plotted in 
Figures~\ref{modelIII_smf} for comparison, are consistent with the
model predictions.  Also, at the low-mass end ($M_*\approx 10^8 h^{-2}
\Msun$) the satellite fraction predicted by Model III is higher than
that predicted by Model II, and eventually overtakes the fraction of
central galaxies of similar masses. This could be checked by
studying galaxy groups in deeper future surveys.

The posterior model parameters obtained for Model III are listed in
Table~\ref{model}.  As one can see, except for $\gamma$, the values of
all the other parameters obtained from Model III are quite similar to
those obtained from Model II. For the new model parameters introduced
in Model III, we have $\gamma'= -4.5$, $z_c=2.2$, $\gamma_a=2.5$ and
$\gamma_b=-0.89$.  The significant difference of $\gamma'$ from zero
implies that a redshift-dependent $\gamma$ is preferred by the data,
and the fact that $\gamma_a$ is much larger than $\gamma_b$ indicates
that the SFR increases with halo mass much faster at
low redshift ($z\ll z_c$) than at $z\gg z_c$.  The value of $z_c$ is
constrained to $2.2 \pm 0.5$ and, interestingly, is very close to the
value expected from the preheating scenarios mentioned above. We will
come back to discuss further the implications of these results.
For both Model II and Model III, the parameters that control 
the star formation in satellites are not well constrained. 
This indicates that the observational constraints used in this
work are not particularly sensitive to the star formation after 
infall of the satellites. As suggested by many other works, 
other observations, such as the clustering \citep{Yang12,Watson13} or 
the quenched fraction of satellites \citep{Wetzel13} could lead to
a better understanding of the star formation history of satellite
galaxies.

\begin{figure*}
\centering
\includegraphics[width=0.5\linewidth,angle=270]{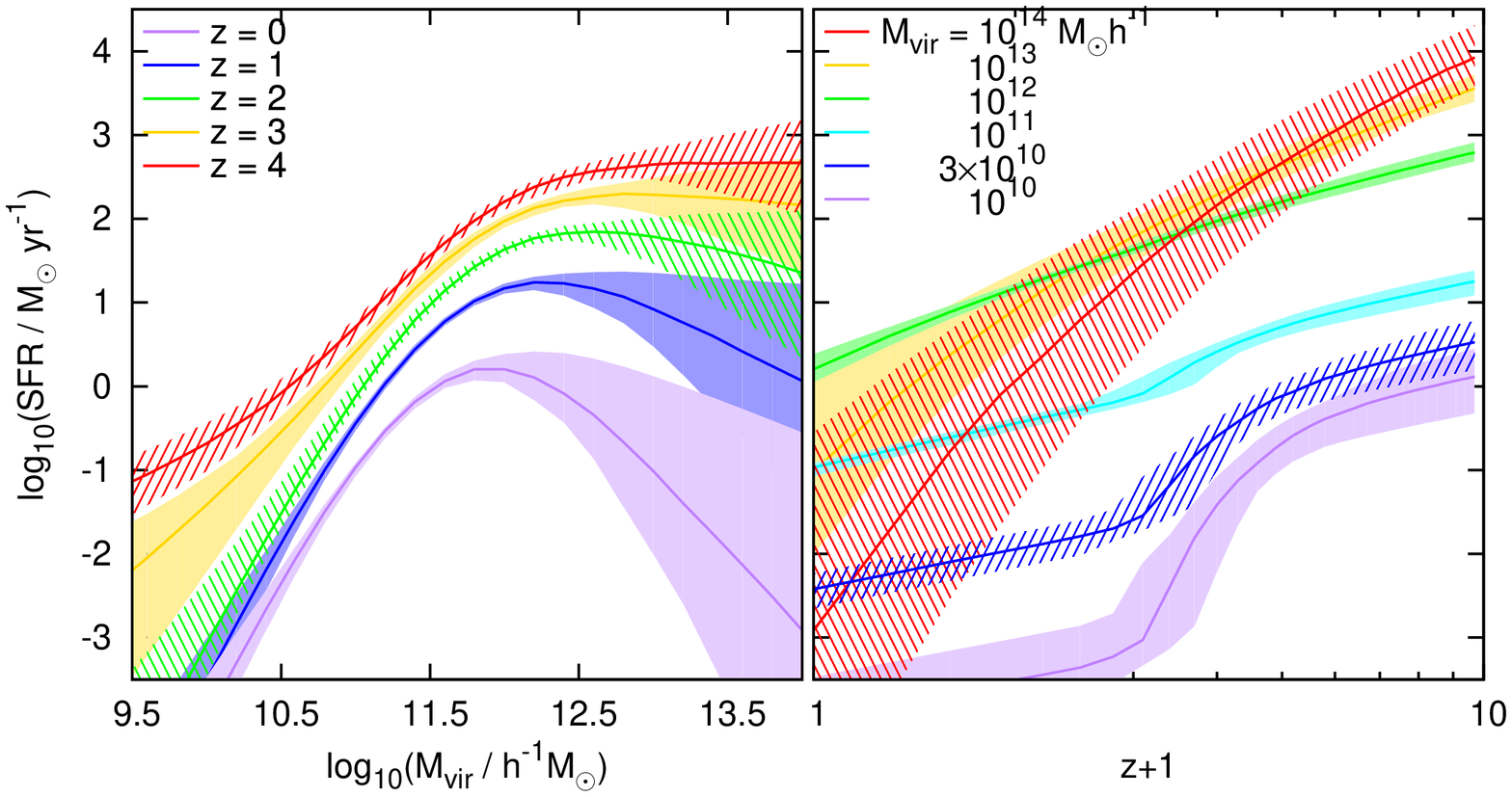}
\caption{The SFR as a function of halo mass (left panel) 
       and redshift (right panel) predicted by Model III.
       The solid lines are the medians of the posterior predictions 
       and the bands are the 95\% credible intervals.}
\label{modelIII_map}
\end{figure*}

Figure~\ref{modelIII_map} shows the model prediction of Model III for
the SFR as a function of halo mass and redshift.  
For halos more massive than $10^{11}{\rm M_{\odot}}$, 
the results are almost identical to those given by Model II. 
For less massive halos, however, Model III predicts a clear
transition at $z_c \approx 2$ from a phase of elevated star formation
at higher $z$ to a phase of reduced star formation at low $z$, 
as can be seen in the right panel of Figure~\ref{modelIII_map}.  
This behaviour owes to the adding of the
cluster galaxy luminosity function data instead of the use of the more
extended Model III. When using only the four SMFs as
constraints, the less restrictive model III is only preferred over
Model II by a factor $e^{0.6} \approx 1.8$, and the posterior for the
star formation rate is similar to that of Model II shown in
Figure\,\ref{modelII_map}. However, when including data constraints
from the cluster luminosity function, the Bayes factor increases to
$e^{26.4}$.

\subsection{Are more general model families necessary?}

As shown above, Model I successfully matches the observed $z\approx 0$
SMF, Model II successfully matches the observed galaxy SMFs over 
the redshift range from $z=0$ to $z=4$, and Model III successfully matches 
not only the observed galaxy SMFs but also the z-band CGLF at $z\approx 0$.  
Perhaps the observational data could be fit even better with a more complex model?
Is the resulting SFR as a function of halo mass and redshift unique 
or are there other models that could match the
observational data equally well but predict a $SFR (M_{\rm vir}, z)$
that is very different from that predicted by the previous models?
To investigate these questions we performed three more inferences.
First, using only the $z\approx 0$ SMF as data constraint for Model II, 
Table~\ref{summary} shows that this data only marginally prefers 
the more complex Model II over Model I and hence Model I is sufficient 
to describe the $z\approx 0$ SMF.  
Next, we performed an inference with Model III using the four
observed galaxy SMFs as data constraints only.  Again,
Table~\ref{summary} shows that these data only modestly prefer
the more complex Model III over Model II, making Model II sufficient
to describe the SMFs over the range $z=0$ to $4$.

Finally, to test the robustness of the Model III predictions using
both the observed galaxy SMFs and also the CGLF, we consider another model family, 
Model IV, which allows even more model parameters to be redshift-dependent.
Specifically, we write
\begin{equation}
   \gamma = 
  \begin{cases}
    \gamma_{\rm a} \left(\frac{z+1}{z_{c}+1}\right)^{\gamma''}   
         \, & \text{if}\, z < z_{c} \\
    (\gamma_{\rm a}-\gamma_{\rm b})
     \left(\frac{z+1}{z_{c}+1}\right)^{\gamma'} + 
     \gamma_{\rm b}     
         \, & \text{otherwise}\,;
  \end{cases} 
\end{equation}
\begin{equation}
    M_{\rm c} = M_{\rm c,0}(z+1)^{\mu}\,;
\end{equation}
\begin{equation}
    \mathcal{R} 
    = \mathcal{R}_{\rm 0}(z+1)^{\rho}\,; 
\end{equation}
\begin{equation}
\kappa={3\over 2}+\kappa'\,,
\end{equation}
with $\gamma''$, $\mu$, $\rho$, $\kappa'$ introduced as four new free
parameters. Model IV reduces to Model III if these four parameters are
set to be zero.

\begin{figure*}
  \centering
  \includegraphics[width=0.8\linewidth,angle=270]{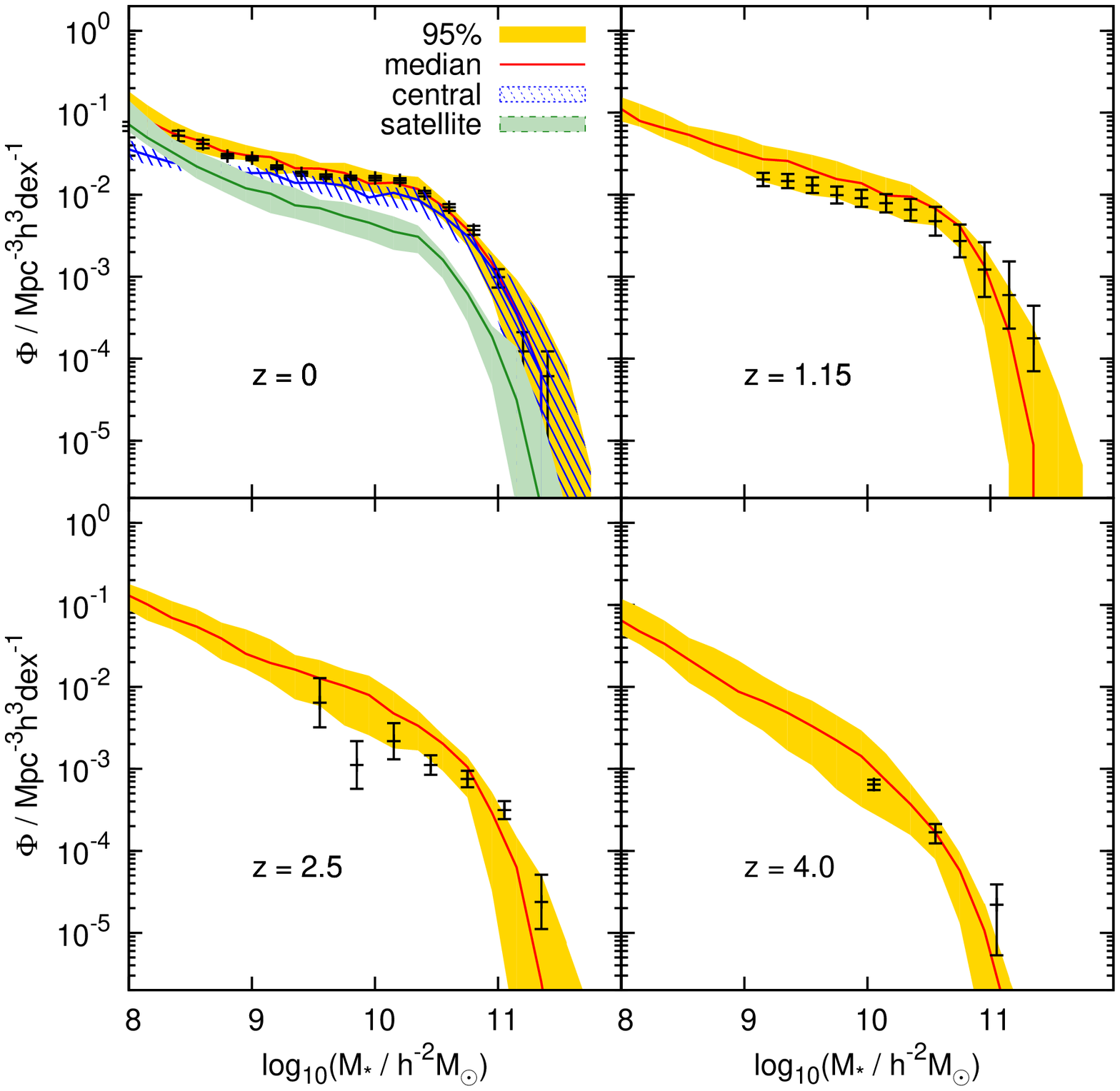}
  \caption{The galaxy SMFs predicted by the posterior of Model IV 
    constrained by both the SMFs and the observed $z$-band CGLF.  
    The yellow bands encompass the 95\% credible
    intervals of the posterior distributions, and the red solid lines
    are the medians. The black data points with error bars are the
    observational constraints.}
\label{modelIV_smf}
\end{figure*}

\begin{figure}
  \centering
  \includegraphics[width=0.9\linewidth,angle=270]{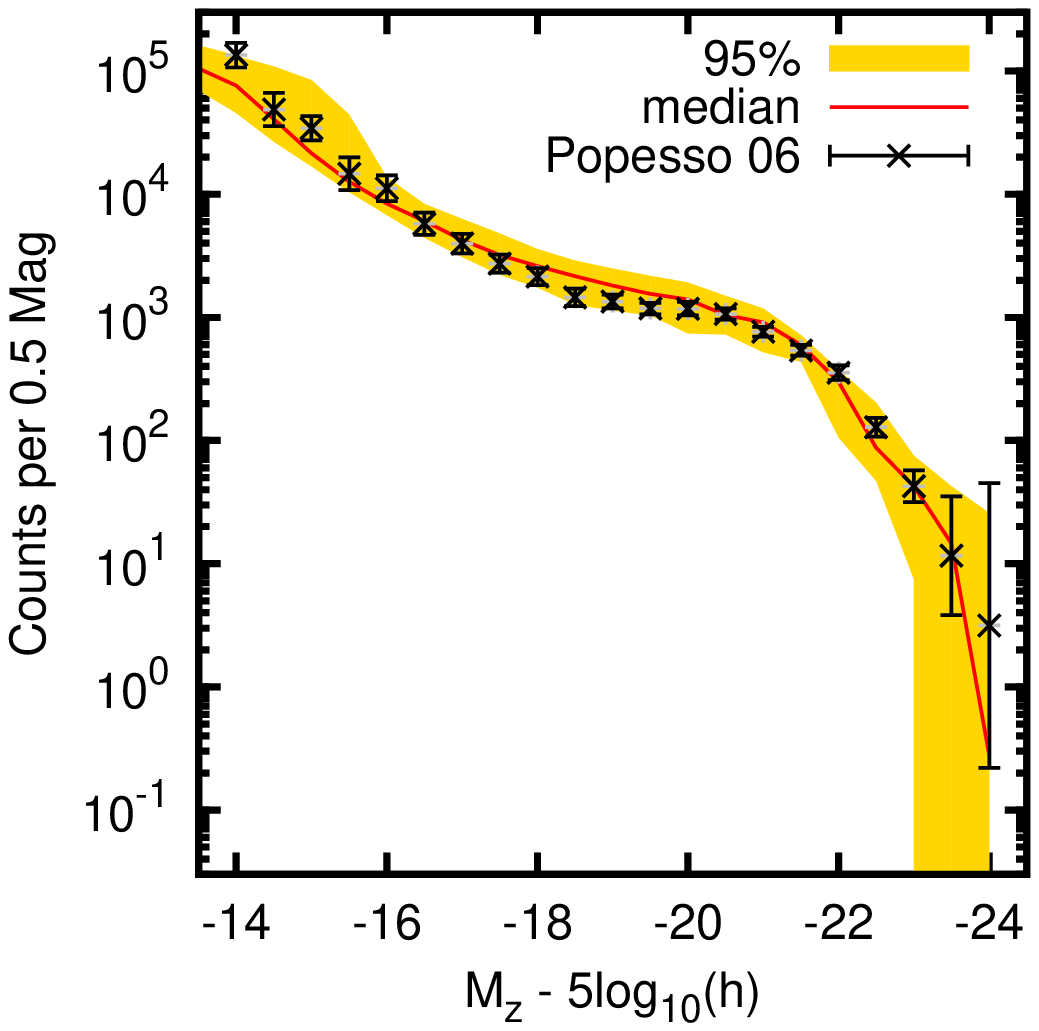}
  \caption{The CGLF predicted by the posterior of Model IV 
    constrained by both the SMFs and the observed $z$-band CGLF.  
    The yellow band encompasses the 95\% credible
    interval of the posterior distribution, and the red solid line is
    the median. The black data points with error bars are the
    observational constraints.}
\label{modelIV_lf}
\end{figure}

\begin{figure*}
\centering
\includegraphics[width=0.5\linewidth,angle=270]{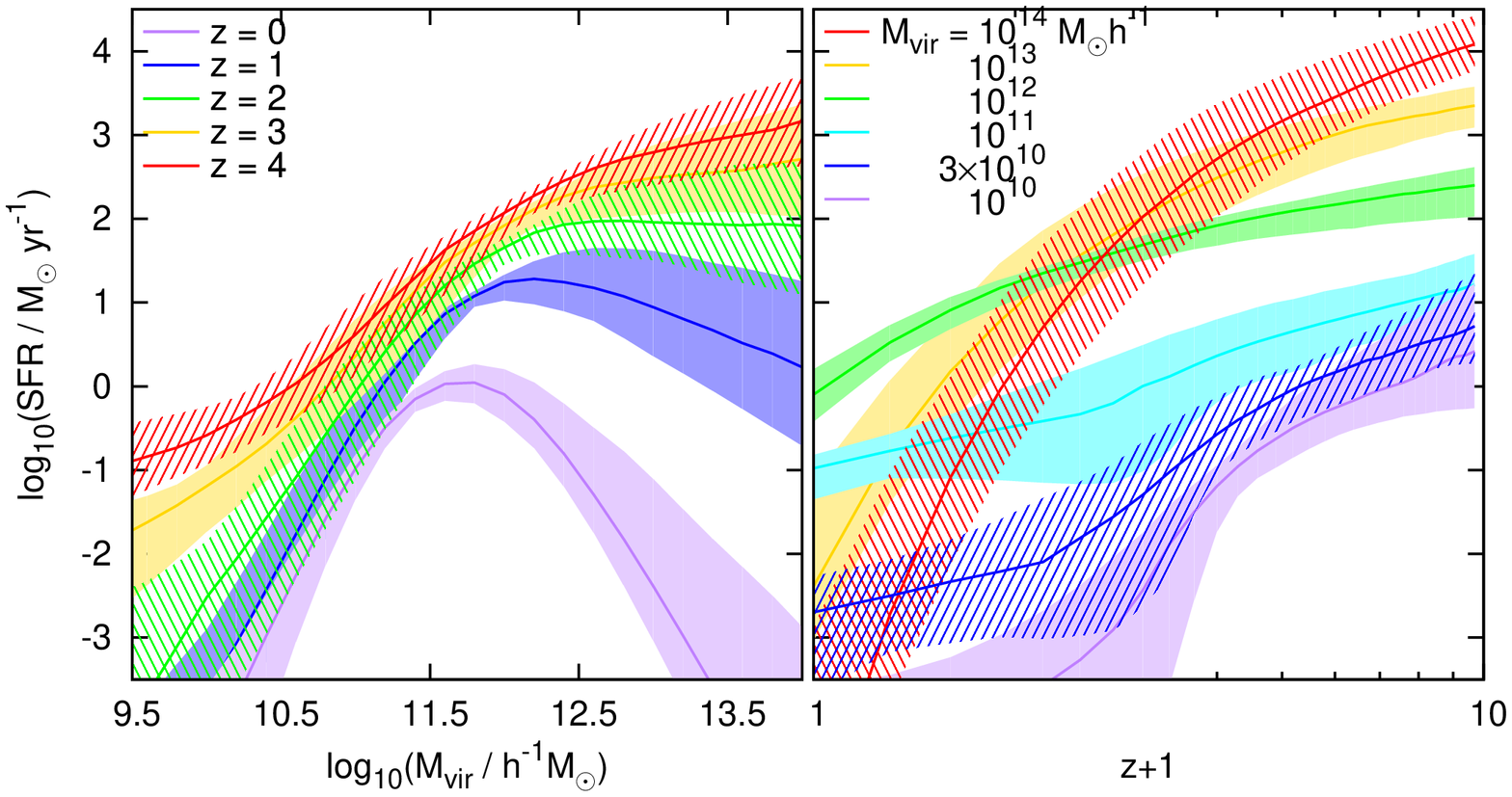}
\caption{The SFR as a function of halo mass (left
  panel) and redshift (right panel) predicted by Model IV.  The solid
  lines are the medians of the posterior prediction and the bands are
  the 95\% credible intervals.}
\label{modelIV_map}
\end{figure*}

In Figures~\ref{modelIV_smf} and \ref{modelIV_lf} we show the
posterior predictions of Model IV for the galaxy SMFs and CGLF respectively,
compared with the corresponding constraining data. As expected, the
larger parameter number Model IV fits the constraining data better. 
In terms of the Bayes Factor, the ratio between Models IV 
and III is $e^{4.9} \approx 134$, which represents a marginally 
significant improvement. An improvement in the fit 
of the $z\approx 0$ SMF near the knee is evident, 
but no other significant improvements are noticeable.  
Furthermore, the SFR as a function of halo mass and redshift 
predicted by Model IV is qualitatively similar
to that predicted by Model III, as shown in Figure~\ref{modelIV_map}. 
The only significant difference is that the
star formation in massive halos with $M_{\rm vir} > 10^{12.5}
\Msunh$ predicted by Model IV increases faster with increasing
redshift than that predicted by Model III.  
Thus, the SFR as a function of halo mass and redshift predicted by Model III
does not seem to owe to the particular parameterisations adopted but
rather reflects requirements of the observational data. Unfortunately,
such tests can never be exhaustive; there is always the possibility
that some other model could match the data constraints and yet give a
different ${\rm SFR}(M_{\rm vir},z)$. In this sense, all our
conclusions and predictions are restricted to the model families that
we actually explore.

Since Model IV and Model III make similar predictions for the star
formation histories for halos with different masses, we will base our
following presentation on Model III.  For reference the posterior
model parameters of Model IV are also listed in Table~\ref{model}.

\section{Model Predictions}

In the following, we use our posterior distributions to make predictions,
both to explore some observational consequences of our models and to
better understand the physics that may give rise to our model. We
present results for both Model II, constrained with just the SMFs 
at different redshifts, and Model III, constrained
using both the SMFs and the z-band CGLF. We feel that presenting Model II's
predictions could still be informative because: 1) it is possible that
the stronger faint-end upturn of the CGLF turns out to be spurious, 
and 2) Model II is similar to the results of other past work
\citep{Yang12, Behroozi12, Behroozi13, Moster13, Mutch13} that also do
not use the CGLF as a constraint.

\subsection{The star formation history and the stellar mass assembly of galaxies}
\label{sec:mah}

\begin{figure*}
 \centering
 \includegraphics[width=0.8\linewidth,angle=270]{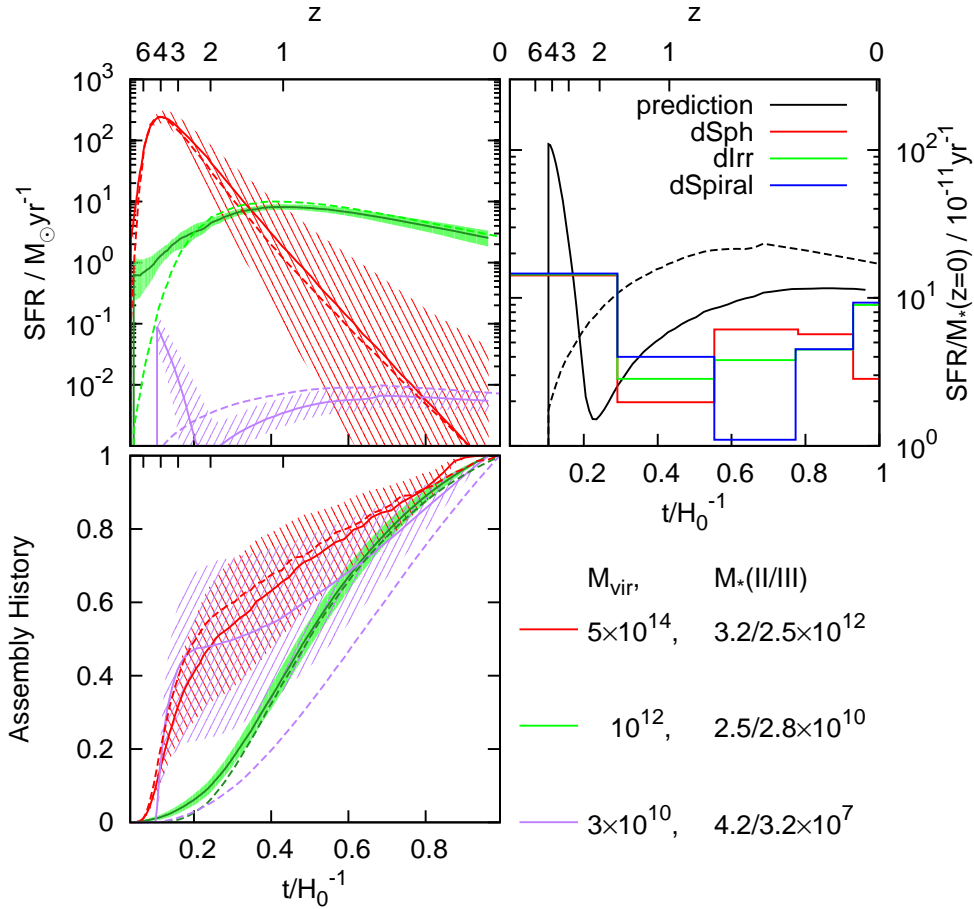}
 \caption{Left panels: the star formation history (upper panel) and
   assembly history (lower panel) of present-day central galaxies in
   halos of different masses. The mass of the host halo and the
   stellar mass of the galaxy (both at the present time) are indicated
   in the lower right panel, both in units of $\Msunh$. The lines are
   the medians and the shaded areas encompass the 95\% ranges of the
   posterior predictions.  Top right panel: the predicted star
   formation history of dwarf central galaxies in halos with masses $\approx
   3\times10^{10}\Msunh$ compared with the measurements of
   \citet{Weisz11} for 60 dwarf galaxies of different types.  
   In all panels, the predictions of Model II are shown by
   dashed lines while the corresponding predictions of Model III are
   shown by solid lines with the same colour.}
 \label{gsfh}
\end{figure*}

Let us first look at the the star formation and assembly histories of
present-day galaxies. Figure~\ref{gsfh} shows the predictions of Model
II (dashed lines) and Model III (solid lines).  Each of the curves
shown is the average over a large number of galaxies with the same
halo mass at $z=0$, as indicated in the lower right panel. For
comparison, the average final stellar mass of the central galaxies for
each halo mass is also indicated in the lower right panel. The top
left panel shows the star formation history, which 
takes into account the contribution of all the progenitors.
The two models yield similar results for galaxies with
present stellar masses $M_{*} > 10^{10}\Msunh$ (halo masses $M_{\rm
vir} \ge 10^{12}\Msunh$). The SFR shows a peak that
shifts from $z\approx 4$ for massive halos ($M_{\rm vir} \sim
10^{15}\Msunh$) to $z\approx 1$ for Milky-Way mass halos ($M_{\rm vir}
\sim 10^{12}\Msunh$).  It declines exponentially after the peak, 
and the rate of decline strongly correlates with halo mass (or galaxy mass), 
from a very fast decline for massive halos to an almost 
constant SFR for halos with $M_{\rm vir} \sim 10^{12}\Msunh$.  
For galaxies in halos with $M_{\rm vir}<10^{12}\Msunh$, the star formation histories
predicted by the two models have different characteristic shapes,
although the predicted final stellar masses are similar. The star
formation histories of dwarf galaxies predicted by Model II show a
rapid increase before becoming almost flat. For Model III, on the
other hand, the average star formation history in low-mass halos is
bimodal: it starts with a high value at $z>z_c\approx 2$, declines
with time to a minimum at $z=z_c$, and then increases to an extended
period of an almost constant star formation. Hence, a
`smoking gun' difference between models II and III is that the
latter predicts a much larger fraction of old stars in (central)
dwarf galaxies than model II.

Indeed, observations of nearby dwarf galaxies indicate that such an
old stellar population is ubiquitous. With the use of deep HST
imaging, individual stars of nearby dwarf galaxies can be resolved,
and the colour-magnitude diagram (CMD) can be constructed to obtain
the detailed star formation histories of these galaxies.  Using
this technique, \citet{Weisz11} investigated 60 dwarf galaxies within
a distance of 4 Mpc, many of which are field galaxies located outside
the Local Group.  These galaxies cover a wide range of morphological
types, including dEs, dIrrs and dSpirals. They found that, on average,
the dwarf galaxies formed $60\%$ of their stars by $z\approx 2$ and
$70\%$ by $z\approx1$, regardless of morphological type.  In the top
right panel of Figure~\ref{gsfh} we compare our model predictions with
the CMD-inferred SFHs obtained by Weisz et al. (2011). While the
predictions of Model III are in qualitative agreement with the
data, Model II predicts an age distribution of stars that is clearly
too skewed towards relatively young stars. 

Finally, the bottom panel shows the assembly history:
the total stellar mass contained in the main progenitor branch
of a halo as a function of time, normalised by the final stellar mass. 
Note that the assembly history includes stellar mass growth both by in-situ 
star formation and by mergers with satellite galaxies.
Although the most massive model galaxy formed most of its
stars as early as $z\approx4$, it is assembled much later;
about half of its final mass is added at $z<2$ (see the red curves). 
On average, Milky-Way mass galaxies experienced a dramatic increase 
in their masses after $z\approx 2$; less than $10\%$ percent of their 
present-day stellar masses were assembled by $z=2$. Thus, 
to identify the progenitors of present-day Milky-Way mass galaxies
at $z\approx 2$, one has to look at galaxies with stellar 
masses $\approx10^9\Msun$. Note that the 95\% range of the 
posterior model prediction is quite broad for both massive and 
low-mass galaxies, suggesting that better observational data are 
required to provide more stringent constraints on the model.   
  
\subsection{The stellar mass - halo mass relation of central galaxies}

\begin{figure*}
\centering
\includegraphics[width=0.8\linewidth,angle=270]{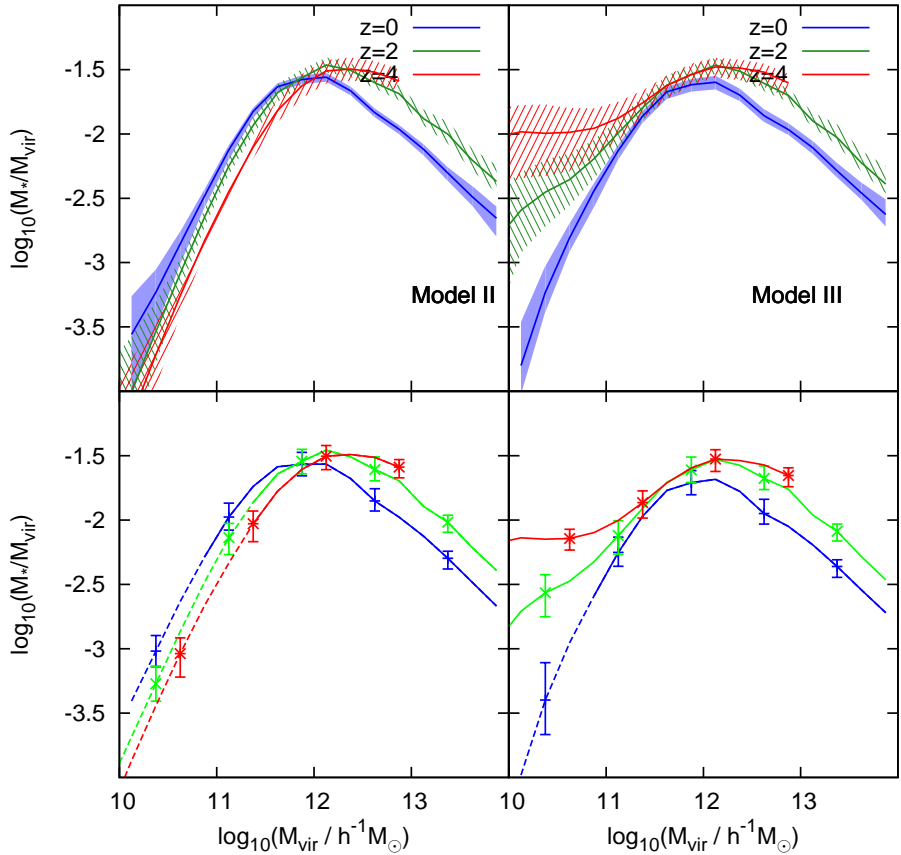}
\caption{The stellar mass to halo mass ratio as a function of halo
  mass for Model II (left panels) and Model III (right panels).  The
  upper panels show the medians of the posterior prediction as well as
  the 95\% inference ranges (bands).  The lower panels show the
  prediction of the best fitting model parameters, and the variance
  among individual halos owing to their different formation histories
  (error bars).}
\label{model_smhm}
\end{figure*}

The left panels in Figure~\ref{model_smhm} show the stellar mass to
halo mass ratio as a function of halo mass at different redshifts as
predicted by Model II.  The dispersion owing to parameter variance
and a variance in the merger histories of the host halos are shown
separately.  These results are similar to those obtained from earlier
investigations \citep[e.g.,][]{Conroy09, Yang12, Behroozi12,
Leauthaud12, Moster13} The ratio shows a broad peak around
$10^{12}\Msunh$ with a gradual shift towards larger halo masses at
higher redshifts. The right panels show the prediction of Model III. 
Compared with the
predictions of Model II, we see a different evolution in halos with
masses below $2\times10^{11}\Msunh$, where the stellar mass to halo
mass ratio is independent of halo mass at high redshifts, owing to an
almost constant star formation efficiency.

\subsection{Star formation rate and halo mass accretion rate}

\begin{figure*}
\centering
\includegraphics[width=0.8\linewidth, angle=270]{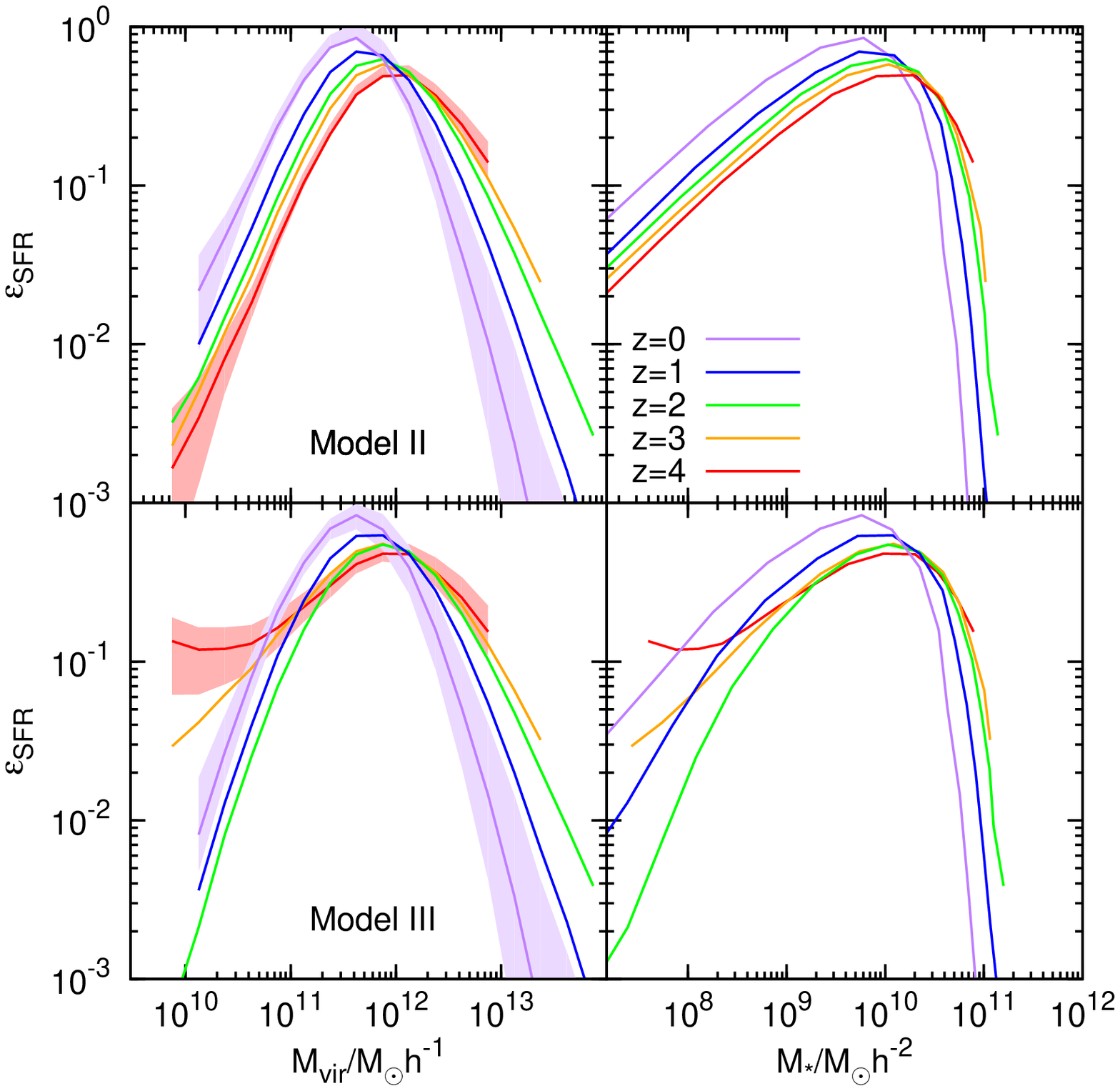}
\caption{ The median star formation efficiency as a function halo mass
  $M_{\rm vir}$ (left panels) and as a function of stellar mass (right
  panels) for halos at different redshifts: $z=0$ (violet curves); 1
  (blue); 2 (green); 3 (orange); and 4 (red).  For clarity, the 95\%
  credible range of the posterior prediction is shown only for $z=0$
  and $z=4$ in the left panels.  The upper panels show the prediction
  of Model II while the lower ones show the prediction of Model III. }
\label{model_eff}
\end{figure*}

One way to understand the star formation history in a halo is to
examine how the SFR correlates with the mass accretion
rate of the host halo. To do this we follow \citet[][]{Behroozi13}
and define a star formation efficiency factor, which is the star
formation rate at a given redshift, ${\dot M}_*(z)$, divided by the
mean mass accretion rate at the same redshift, ${\dot M}_{\rm
  vir}(z)$, times the universal baryon fraction $f_B$ ( we refer to
$f_B {\dot M}_{\rm vir}(z)$ as the baryonic accretion rate):
\begin{equation}
\epsilon_{\rm SFR}(z)
\equiv {{\dot M}_\star(z)\over f_B{\dot M}_{\rm vir}(z)}\,.
\end{equation}
The left two panels of Figure~\ref{model_eff} show the star formation
efficiency as a function of $M_{\rm vir}$, and the right two panels
show the same star formation efficiency as a function of $M_*$.  The
predictions of Model II (upper two panels) and Model III (lower two
panels) are almost identical for host haloes with masses above
$10^{12}\Msunh$ (stellar masses above $10^{9}\Msunh$).  For low-mass
halos, the two models behave quite differently.

Let us first look at the results for Model II. 
The star formation efficiency at $z\approx 4$ is strongly peaked at $M_{\rm vir}\approx
10^{12}\Msunh$, with a peak value $\epsilon_{\rm SFR}\approx 1/2$. The
position and height of the peak depend mildly on redshift; at
$z\approx 0$ it shifts to $M_{\rm vir}\approx 4\times 10^{11}\Msunh$,
with a peak value $\epsilon_{\rm SFR}\approx 0.8$. 
$\epsilon_{\rm SFR}$ increases (decreases) with halo mass 
as a steep power-law at the low (high) mass end.
These results are similar to those obtained by 
\citealt{Bouche10, Behroozi13, Tacchella13}.
In the toy model proposed by \citet{Bouche10}, the SFR
in halos with masses between $3 \times 10^{11} \Msunh$ and $2 \times
10^{12} \Msunh$ follows the baryonic accretion rate, and is completely
quenched in halos outside this range.
Those studies also imply that the star formation efficiency
shows no dependence on redshift. Our results, on the other hand, 
reveal a slow but steady increase of $\epsilon_{\rm SFR}$ near the peak with
decreasing redshift, which owes to a decrease in the halo
assembly rate, ${\dot M}_{\rm vir}$, of halos with masses $\la
10^{12}\Msunh$ at low $z$. While in halos that are much more massive
than $10^{12}\Msunh$, the star formation efficiency increases with
redshift. Such evolution with redshift agrees with the results
from \citealt{Yang13, Bethermin13}.

Model III predicts a different behaviour for halos with masses
$<10^{11}\Msunh$ than Model II or the simple models proposed by
\citet{Bouche10, Behroozi13, Yang13, Bethermin13}, which implies a simple but
strong quenching of star formation. Beyond $z_{\rm c}$ at $z\approx
4$, the SFR is roughly about $1/10$ of the baryonic
accretion rate, and is independent of the host halo mass. The star
formation is quenched abruptly at $z\approx z_{\rm c}$ if the halo is
much smaller than $10^{11}\Msunh$ and makes a mild recovery at
$z\approx 0$.

\subsection{The cosmic star formation history}

\begin{figure*}
 \centering \includegraphics[width=0.8\linewidth,angle=270]{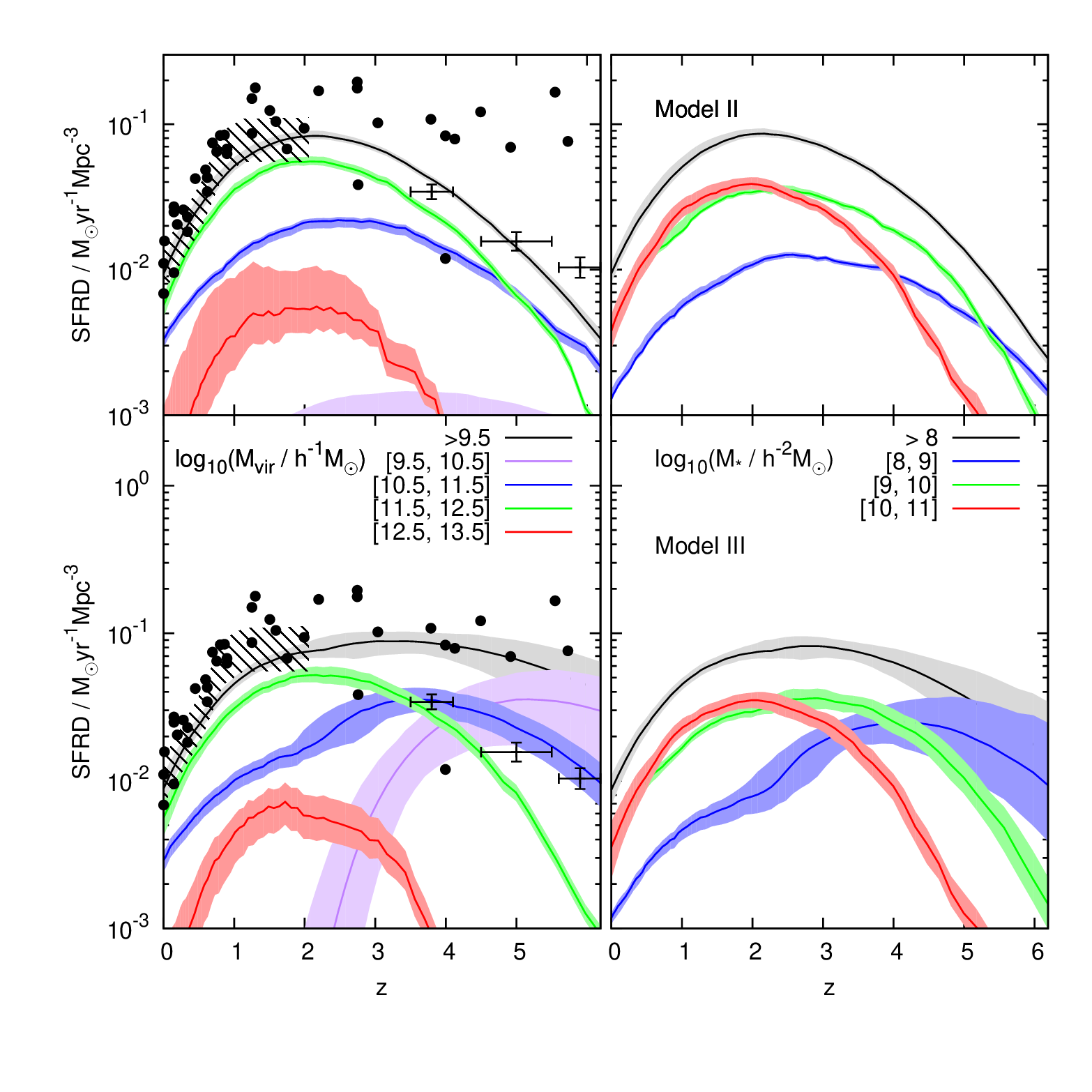}
   \caption{The cosmic star formation rate density (SFRD) as a
     function of redshift predicted by Model II (upper panels) and
     Model III (lower panels).  The solid lines and the bands are the
     medians and 95\% credible intervals, respectively. The total star
     formation rate density is shown as the grey band. The data points
     are taken from \citealt{Hopkins06} (black dots) and Bouwens et
     al. (2012) (the shaded area and the data points with error
     bars). In the left panel we plot the contributions by halos of
     different masses separately, while in the right panel we plot the
     contributions by galaxies of different stellar masses.}
\label{model_sfd}
\end{figure*}

We compare the predicted star formation rate density (SFRD) of the
universe with observations.  Results are shown for Model II (upper
panels) and Model III (lower panels) in Figure~\ref{model_sfd}.  The
total SFRD is decomposed into contributions by halos of different
masses (left panels) and into contributions by galaxies of different
stellar masses (right panels). 
The observations shown in the figure are compiled by
\citet{Hopkins06} and \citet{Bouwens12}.  The data points shown here
are produced using a Chabrier IMF.  At $z \le 2$ the observational
results from different sources are consistent with each other, and so
are the predictions of both Model II and Model III.  All of them show
a fast decline, by an order of magnitude, towards low-$z$ beginning at
$z=2$. In both Model II and Model III, the predicted total SFRD in
this redshift range owes mainly to star formation in halos with masses between
$3\times10^{11} \Msunh$ and $3\times10^{12}\Msunh$, about Milky Way
mass.  As one can see from Figures~\ref{modelIII_map} and
\ref{modelII_map}, the SFR in such halos is proportional to $(1+z)^{2.3}$. 
However, at $z>3$ the two models behave differently. 
Model II predicts a rapid decline in the total SFRD
towards high $z$, as the abundance of $10^{12}\Msunh$ halos decreases
towards higher redshift while in dwarf halos, which are abundant at
high-$z$, the star formation is strongly suppressed.  The SFRD
at $z>3$ predicted by Model III is substantially higher, mainly
because the SFR in low-mass halos
($<3\times10^{11}\Msunh$) or dwarf galaxies ($<10^{9}\Msunhh$) is
boosted at $z>z_c$ in this model.  This difference provides an
observational test to distinguish these two models.  Unfortunately,
current observational results of the SFRD are still quite uncertain. 
A simple comparison between observations and our model predictions 
shows that the data of \citet{Hopkins06} favours Model III but that 
of \citet{Bouwens12} favours Model II. 
The difference in the data owes to the uncertainty in dust corrections 
\citep[e.g.][]{Reddy09}. Also, corrections for sample incompleteness
contain large uncertainties. One derives the high-$z$ SFRD by integrating the
observed UV luminosity function.  Unfortunately, the faint-end
behaviour of the high-$z$ luminosity functions is usually poorly
constrained, and the exact value of the assumed faint-end slope can
affect the derived SFRD significantly. It may still be  
that the dust correction used by \citet{Bouwens12} is 
correct, but that one underestimates the total cosmic SFR at 
high-$z$ because one misses a large number of low-mass galaxies. 

\subsection{The star formation rate function}

\begin{figure*}
   \centering
   \includegraphics[width=0.8\linewidth,angle=270]{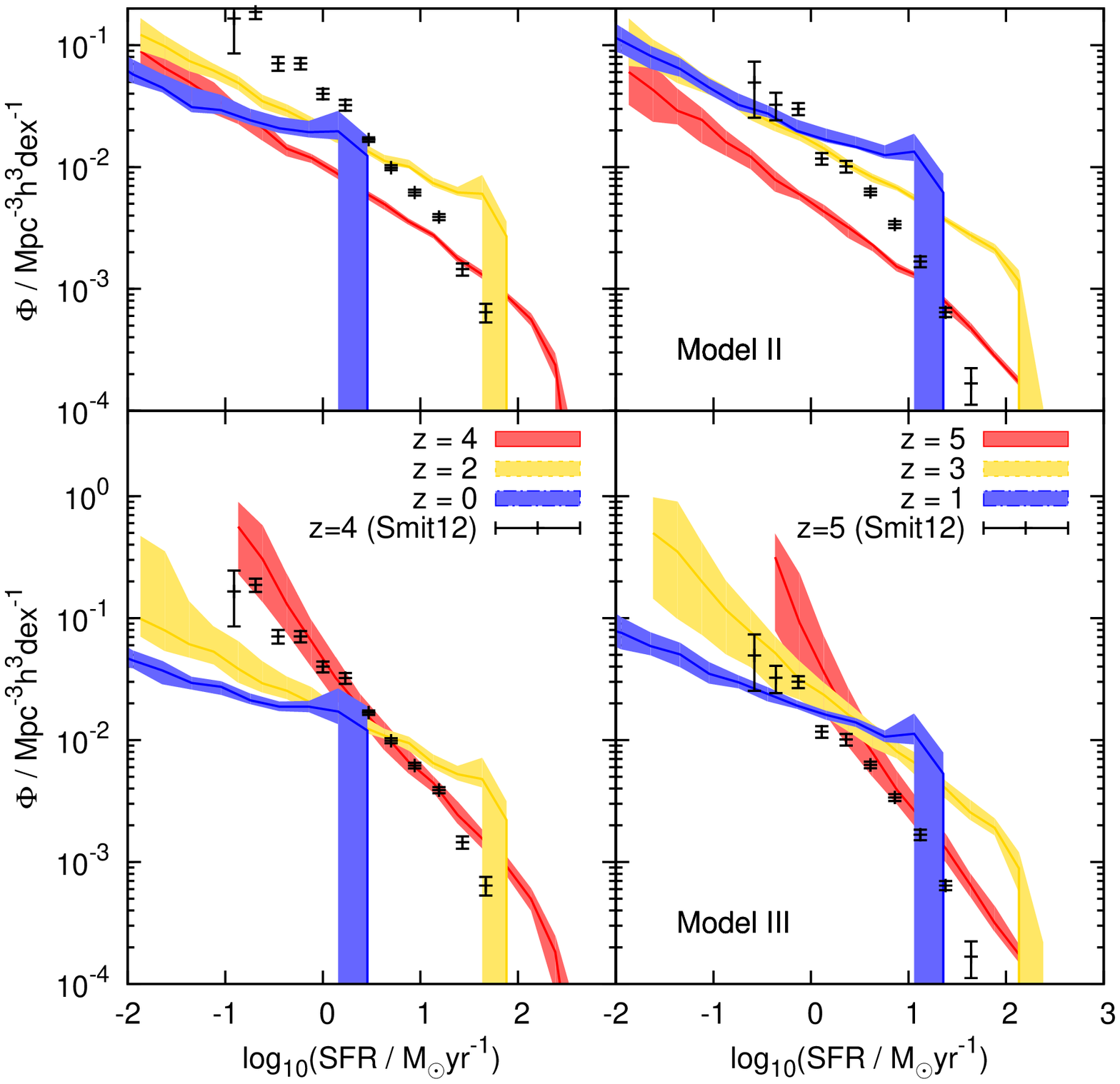}
   \caption{The star formation rate functions at different redshifts
     predicted by Model II (upper panels) and Model III (lower
     panels).  The solid lines are the medians and the bands encompass
     the 95\% credible intervals. The data points are the
     observational results from \citet{Smit12}.}
\label{model_sfrfunc}
\end{figure*}

In Figure~\ref{model_sfrfunc} we show the predicted star formation
rate functions for galaxies at different redshifts. Note that
none of these functions can be well fit with a Schechter function.
For $z<3$, there is a sharp cutoff following a bump at the high-SFR
end. This owes to the existence of peaks in the star formation
rate-halo mass relations (see Fig.\,9). However, this feature should
not be taken too seriously because what we show here is based on the
average SFR - halo mass relation, ignoring any
dispersion in the relation.  Despite this, the characteristic star
formation rate clearly decreases with decreasing redshift, by a factor
of almost 100 from $z=4$ to $z=0$.  Assuming that the faint part of
the distribution can be fit by a power law, the power law index
predicted by Model III changes significantly from roughly $-2.0$ at
$z=4$ to roughly $-1.2$ at $z=0$. For Model II the change in the faint
end slope is much more moderate, from roughly $-1.5$ at $z=4$ to
$-1.2$ at $z=0$.  For comparison we also show the SFR function derived
by \citet{Smit12} from the UV luminosity function of galaxies.  We see
that Model II significantly underpredicts the number density of
galaxies at the low-SFR end, while Model III matches the data much
better. Note that the observed SFR functions at the high-SFR ends are
lower than the predictions of both Model II and Model III.  One
possible reason for this discrepancy is that the highest SFR galaxies
are dusty and could be missed in UV observations.

\subsection{The specific star formation rate}

\begin{figure*}
 \centering
 \includegraphics[width=0.8\linewidth,angle=270]{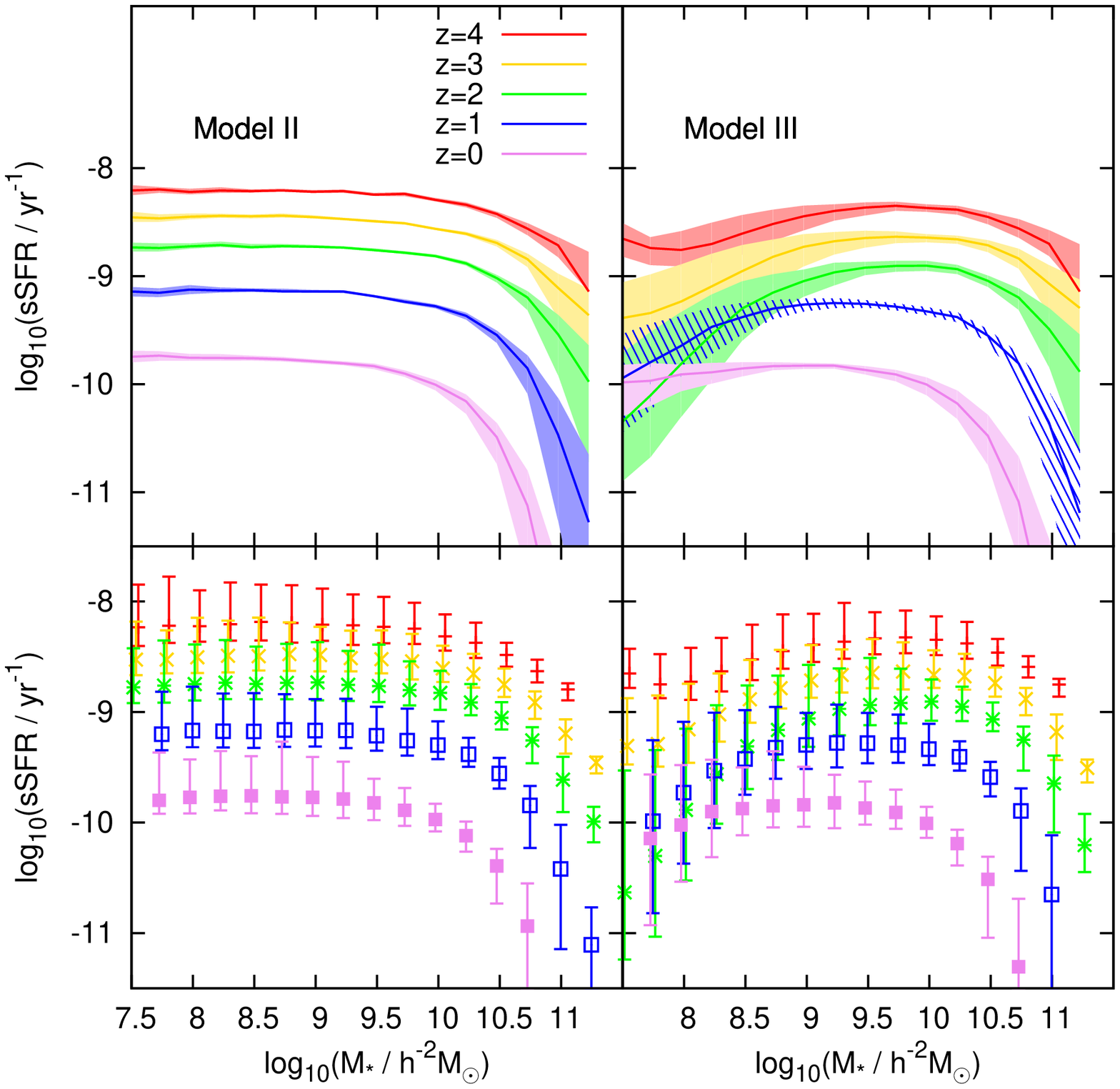}
 \caption{The specific star formation rate (sSFR) versus stellar mass
   at different redshifts in Model II (left panels) and Model III
   (right panels).  The upper panels show the medians of the posterior
   predictions as well as their 96\% ranges (bands).  The lower panels
   show the predictions of the best fitting model parameters, and the
   variance among individual halos owing to their different merger
   histories (error bars).}
\label{model_ssfr}
\end{figure*} 

In Figure~\ref{model_ssfr}, we show the specific star formation rate
(sSFR; defined to be the SFR divided by the stellar mass) as a
function of stellar mass at different redshifts. Here again we compare
between the predictions of Model II (left panels) and Model III (right
panels).  For a given stellar mass, the sSFR increases with
redshift. On the other hand, for a given redshift, the sSFR declines
rapidly with galaxy mass as the mass goes beyond a critical mass,
which increases from $\approx 10^{10}\Msunhh$ at $z=0$ to $\approx
5\times10^{10} \Msunhh$ at $z=4$.  For galaxies between $10^9 \Msunhh$
and the critical mass, the sSFR is almost independent of stellar mass,
in qualitative agreement with the observations
  \citep[e.g.][]{Daddi07,Noeske07a}.  Model II and Model III differ
in their predictions for low-mass galaxies with stellar masses $<10^9
\Msunhh$.  For Model II, the weak correlation between sSFR and stellar
mass extends all the way down to such galaxies. For Model III, however, the
correlation is much more complicated: the sSFR and stellar mass show
no significant correlation at $z=0$, show a strong positive
correlation between $z=1$ and $z=2$, and show a weak positive
correlation at higher redshifts.  For low-mass galaxies at high $z$,
Model III predicts a lower sSFR compared to Model II, because these
galaxies in Model III form their stars earlier than galaxies of the
same mass in Model II. In Model III the SFR in
low-mass galaxies drops dramatically at the critical redshift
$z_{c}\approx 2$ after a significant amount of stars have already
formed at higher $z$, making the sSFR in dwarf galaxies much lower
than that in more massive galaxies at the same epoch.  At $z\approx 0$
the sSFR in dwarf galaxies catches up with those in more massive
galaxies, because of the growth of their halos and because of the
strong mass dependence of the SFR at the low-mass end
at low $z$.

Observations indicate that there is a negative correlation between
sSFR and stellar mass for dwarf star forming galaxies
\citep{Noeske07a}, a trend that appears contrary to the
predictions of Model III shown in the right panel.  However, the
existence of such a correlation in the observational data is still
uncertain owing to sample incompleteness.  Indeed, empirical models
based on such correlations \citep{Noeske07b, Leitner12} suggest that
low-mass galaxies form most of their stars in the past few billion
years, in apparent contradiction with the star formation histories
inferred directly from the colour-magnitude diagrams of stars
\citep[e.g.][]{Weisz11}, which seem to support Model III (see
\S~\ref{sec:mah}) More data on the detailed SFHs of isolated dwarf
galaxies is required to better discriminate between these
different models. Note that SED modelling is used to infer star formation
rates and stellar masses. If these SED models do not include
the `bimodal' SFHs, such as those predicted by model III, significant
systematic errors in the inferred $\dot{M}_\star$ and $M_\star$ of
dwarf galaxies can occur. It is, therefore, important to check the
magnitudes of such systematic effects.  
  
\subsection{The conditional stellar mass functions}

\begin{figure*}
   \centering
   \includegraphics[width=0.8\linewidth,angle=270]{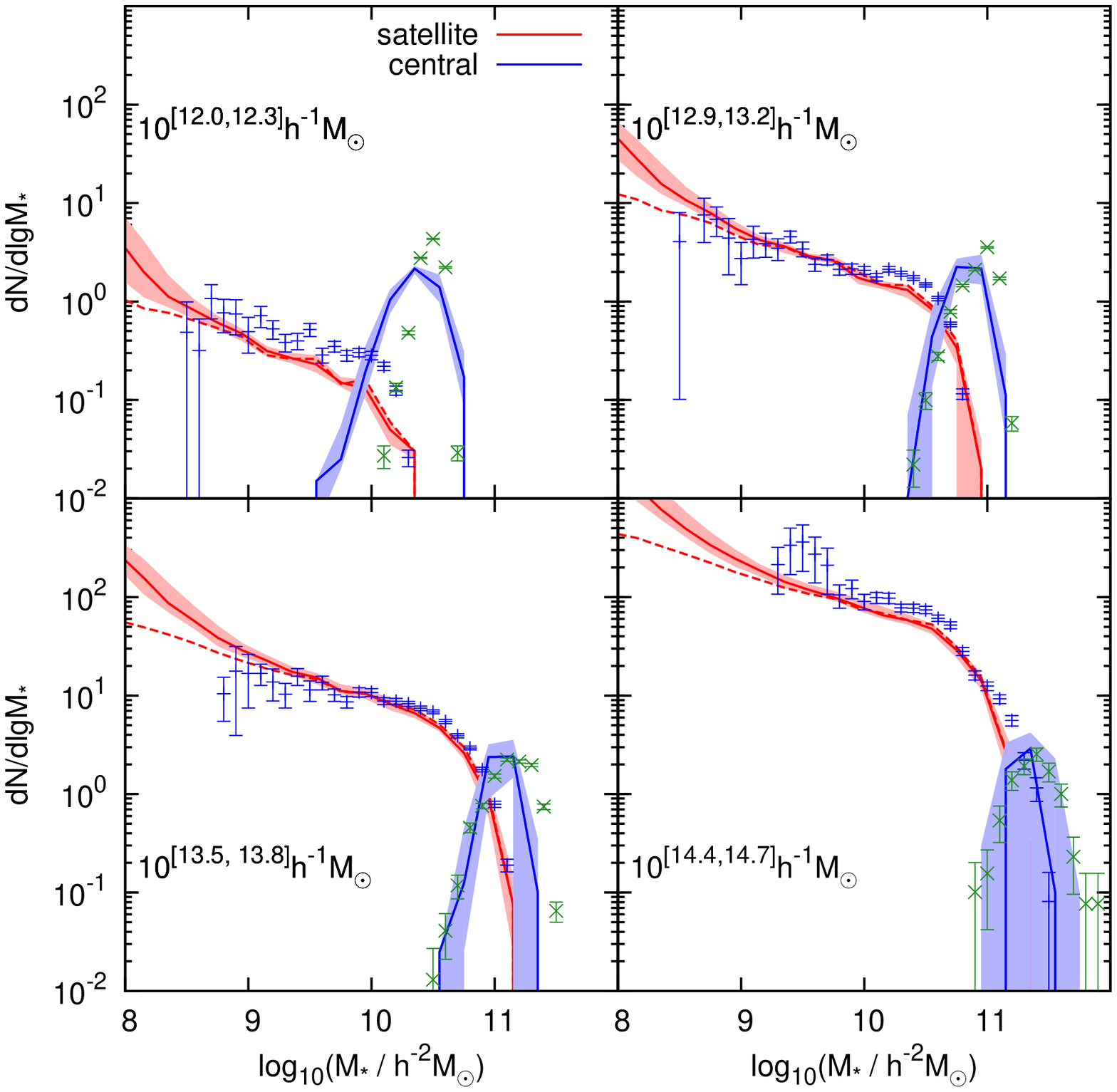}
   \caption{The conditional stellar mass functions for four different
     halo mass bins as indicated. The solid lines with bands are the
     predictions of Model III, while the dashed lines are the
     predictions of Model II.  The distribution of centrals and
     satellites are in blue and red, respectively.  The data points
     are from \citet{Yang08}.}
   \label{model_cond}
\end{figure*}

We also make predictions for the Conditional Stellar Mass Function
(hereafter CSMF), which is the SMF of galaxies
hosted by halos of a given mass. We show them in
Figure~\ref{model_cond} together with the observational results of
\citet{Yang08}. Owing to the detection limits, the CSMF
below $10^9 \Msunhh$ is either noisy or unavailable.  
The predictions of both Model II and Model III above the detection limit are
consistent with the observational data. (Note that a quantitative
comparison requires the prediction be convolved with the effects of the
group finder \citep{Reddick13}, which is beyond the scope of this paper.)
Compared with Model II, Model III predicts more dwarf galaxies, 
with masses below the current detection limit, 
not only in massive clusters but also in low-mass groups. 
One expects this behaviour because in Model III dwarf galaxies
in massive clusters are fossils of a relative global enhancement of
star formation activity in dwarf halos in the high-$z$ Universe. This
boosted star formation at high-$z$ also leaves an imprint in
present-day galaxy systems of lower halo masses, not just in rich
clusters.

\section{Summary and Discussion}

In this paper, we use the observed SMFs of galaxies
in the redshift range from $z\approx 0$ to $z\approx 4$ and the
luminosity function of cluster galaxies at $z\approx 0$ to constrain
the star formation histories of galaxies hosted by dark matter halos
of different masses.  To this end, we parametrise the star formation
rate as a function of halo mass and redshift using piecewise power
laws. We combine this empirical model for star formation with halo
merger trees to follow the evolution of the stellar masses of galaxies
and to make model predictions to be compared with the data
constraints. We use the MULTINEST method developed by \citet{Feroz09}
to obtain the posterior distribution of the model parameters and the
marginal likelihood.  A series of nested model families with
increasing complexity are explored to understand how the model
parameters are constrained by the different observational data
sets. We use the posterior model parameter distributions to make model
predictions for a number of properties of the galaxy population and
compare these results with available observations. Our main results
can be summarised as follows:

\begin{itemize}

\item To match the observed SMFs at different
  redshifts, the SFR in central galaxies residing
  in haloes with masses above $10^{12}\Msunh$ has to be boosted at
  high redshift relative to the increase that arises naturally from
  the fact that the dynamical time scale is shorter at higher $z$.

\item To reproduce the faint end of the cluster and field galaxy
  luminosity functions ($M_{z}-5\log_{10}(h) > -18$) simultaneously, we
  require a characteristic redshift $z_c \approx 2$ above which the
  SFR in low mass halos with masses $<10^{11}\Msunh$
  must be enhanced relative to that at lower $z$. This
  enhancement is also supported by the fact that isolated dwarf
  galaxies seem to be dominated by old stellar populations 
  (\citet{Weisz11}), and by the observed star
  formation rate functions at high redshift \citep{Smit12}.

\end{itemize}

  Our model (Model III) that successfully matches all these observations 
makes the following predictions:

\begin{enumerate}

\item 
  The star formation efficiency, that is the SFR divided 
  by the baryonic mass accretion rate of the host halo, peaks in halos
  with masses between $3\times10^{11}\Msunh$ and $10^{12}\Msunh$.
  In lower mass halos, the star formation efficiency is about $1/10$ 
  at $z>z_c$ and is strongly quenched at lower $z$ and roughly scales
  as $M_{\rm vir}^{3/2}$. While in higher mass
  halos, the star formation tends to be quenched and the quenching is
  stronger with decreasing redshift.

\item The average star formation histories for the central galaxies of
  halos with masses $M_{\rm vir} > 10^{12}\Msunh$ are peaked, with the
  peak redshift shifting from $z\approx 4$ for present-day cluster halos
  with $M_{\rm vir}\sim 10^{15}\Msunh$ to $z \approx 1$ for present-day
  Milky Way mass halos ($M_{\rm vir} \sim 10^{12} \Msunh$).

\item The average star formation history in low mass halos with
  $M_{\rm vir} < 10^{12}\Msunh$ is` bimodal': it starts with a high
  value at $z > z_c$, declines with time to a minimum at $z=z_c$, and
  then increases before it reaches an extended period of roughly
  constant SFR.  Our model, therefore, predicts the
  existence of an old stellar population formed at $z>z_c$ in dwarf
  galaxies, consistent with the results obtained from direct
  observations of the stellar populations in such galaxies
  \citep{Weisz11}.

\item Central galaxies of massive clusters formed most of their stars
  as early as $z \approx 4$, but on average about half of their final
  mass is assembled at $z<2$. On the other hand, Milky-Way mass
  galaxies experienced dramatic increases in their stellar masses
  after $z\approx 2$, and less than 10\% ($\sim10^9\Msun$) of their
  present-day stellar mass was assembled by $z=2$.

\item The stellar mass to halo mass ratio, $M_*/M_{\rm vir}$, for
  central galaxies peaks at a halo mass of $\approx 10^{12}\Msunh$ with
  a value of $\approx 1/30$, quite independent of redshift.  This is in
  excellent agreement with a wide range of observational constraints
  \citep[see e.g.,][and references therein]{Behroozi10, Dutton10}.

\item For halos with masses below $2\times 10^{11}\Msunh$ our model
  predicts that $M_*/M_{\rm vir} \approx 1/100$ at $z = 4$ quite
  independent of halo mass, but this ratio decreases rapidly with
  decreasing halo mass at $z = 0$.

\item The low-mass end slopes of the SMF and the
  star formation rate function steepen toward high redshift.  Central
  galaxies dominate the present-day SMF at $M_*>
  10^{9}\Msunhh$ but satellite galaxies begin to dominate at $M_*<
  10^{8}\Msunhh$. 

\item Halos with $M_{\rm vir} \sim 10^{12}\Msunh$, hosting
  centrals with $M_\star \sim 10^{10}\Msunhh$, dominate the 
  SFRD of the Universe at $z< 3$ while at higher $z$
  star formation in lower mass halos takes over.  Star formation in
  halos more massive than $10^{12.5}\Msunh$ never significantly
  contribute to the total SFRD.
\end{enumerate}

Our findings have important implications for the physical processes
that regulate star formation and feedback.  In general the star
formation rate in a halo depends on the amounts of cold gas that can
be accreted into the halo centre, and on the time scale with which the
cold gas converts into stars. In current theories of galaxy
formation the amount of cold gas in a halo is determined by radiative
cooling and feedback effects. \\

It is well known that radiative cooling
introduces a characteristic halo mass, $M_{\rm cool}\approx
6\times10^{11}\Msun$, which separates cooling limited `hot mode' and
free-fall limited `cold mode' in the accretion of cold gas into
galaxies \citep{Birnboim03, Keres05, Keres09}.  For halos below this
characteristic mass, gas is never heated during accretion and so the
amount of cold gas is limited by the free fall time of the gas. For
halos with larger masses, on the other hand, the accreted gas first
heats by accretion shocks and then cools radiatively
before it sinks into the central galaxy, so that cold gas accretion by
the central galaxy is limited by the radiative cooling time scale.
The characteristic mass scales we find in the star formation
efficiency shown in Figs.\,\ref{model_smhm} and \ref{model_eff} are
very similar to $M_{\rm cool}$, suggesting that radiative cooling
plays an important role in star formation.  Furthermore, since the
cooling time scale decreases faster than the free-fall time scale as
redshift increases \cite[see \S8.4 in][]{Mo10}, cooling may also have
played a role in the enhanced SFR in massive halos at
high $z$ (see \S\ref{ssec_ModelII}).  

However, radiative cooling alone cannot explain why the 
star formation efficiency in lower mass halos is suppressed.  
Even for massive halos, numerical simulations have
shown that the suppression in radiative cooling at low $z$ is not
sufficient to explain the observed low SFRs, and some
heating sources are needed to quench the star formation in massive
galaxies at low $z$.  One popular mechanism is AGN feedback.
Observations show that AGN activity peaks at $z\approx 2$ and declines
towards both higher and lower redshift (e.g. Hopkins et al. 2007),
indicating that super-massive black holes may have already formed in
massive galaxies by $z\approx 2$. Thus, the quenching of star formation
in massive galaxies at $z<2$ may owe to a combination of effective AGN
feedback (e.g. in low-accretion radio mode) and inefficient radiative
cooling owing to the reduced gas density. Similarly, the high star
formation rate in high mass halos at high $z$ may arise from an
increased radiative cooling efficiency combined with reduced AGN
feedback owing to the reduced number of super-massive black holes that
have formed or the presence of cold, filamentary accretion
in these massive halos at high redshift \citep{Keres09}. 
Our results for the star formation in massive galaxies
are in quantitative agreement with these expectations, and a detailed
comparison between these empirical results and theoretical predictions
will provide important insights into the underlying physical
processes. \\

Our results for the star formation in low-mass halos poses a number of
challenges to standard theory. First, since radiative cooling is
expected to be effective at all redshifts in low-mass halos, some
feedback processes must be invoked to suppress the star formation
efficiency in these halos. A popular assumption is that galactic winds
driven by supernova explosions may suppress the star formation in such
halos. In many models considered thus far, the mass loading factor,
which is defined to be the mass loss rate through winds divided by the
SFR, is assumed to be some power of the circular
velocity of the host halo. In contrast, our results indicate that the
star formation efficiency in dwarf halos at $z\approx 4$ is about
$10\%$ of the baryonic accretion rate independent of halo mass. This
suggests that the mass loading of the wind is independent of the halo
mass at high redshift, so that a constant fraction of the accreted gas
is converted into stars and the rest is driven out of the halo as
galactic winds.  

Furthermore, the existence of a transition at
$z=z_c\approx 2$ separating active from quiescent star forming
phases, which is required to explain the faint-end upturn in the
CGLF, is not expected in the conventional supernova feedback model. 
Instead, our results lend support to a scenario in which the IGM is
preheated at $z\approx 2$ and hence the accretion of baryons into low
mass halos is delayed until they become sufficiently massive to allow
significant accretion from the preheated IGM to form stars at lower
redshift.  The exact mechanism for preheating is still unclear.
Possibilities that have been proposed include the formation of
pancakes \citep{Mo05}, an episode of starburst and AGN activity at
$z\ga 2$ \citep{Mo02}, and heating by high-energy gamma rays generated
by blasars \citep{Chang11}. In all these preheating scenarios,
preheating is expected to be at $z\approx 2$, in excellent agreement
with the value of $z_c$ that we find. The preheated entropy of the
IGM is about a few times $10\,{\rm KeV\,cm^2}$, similar to what is
required to match the observed luminosity function and HI mass
functions of present-day galaxies (Lu et al. 2013). \\

The finding that Milky-Way mass galaxies experienced a
significant increase in their stellar masses after $z\approx 2$
contrasts with the fact that the dark halos of these
galaxies assembled their masses at a much slower pace at $z <2$
\citep{Zhao09}. More specifically, the progenitors of present-day
Milky-Way mass galaxies at $z\approx 2$ on average have a stellar mass
that is about $1/15$ of the final stellar mass, while the average halo
mass at the same redshift is about 1/4 of the final halo
mass. Apparently, the star formation in such galaxies is detached
from and delayed relative to the halo accretion. This contrasts
with numerical simulations that predict that the SFR traces the
accretion rate \citep[e.g.][]{Dave12} at late times when the circular
velocity of Milky-Way mass halos changes slowly with time
\citep{ZhaoMoJing03}. Apparently, some process must have delayed the
cold gas accretion relative to the halo mass accretion in such
halos. As discussed above, preheating may operate in this
way. Alternatively, a large fraction of the accreted cold gas may be
ejected and take a relatively long time to reaccrete onto the galaxy
to feed star formation \citep{Oppenheimer10}. Clearly, one needs a
detailed analysis to see if such processes can produce the star
formation and assembly histories of Milky-Way type galaxies that we
obtain here.

\section*{Acknowledgements}

The authors acknowledge P. Popesso for providing the data
of cluster galaxy luminosity functions and F. Feroz for providing
the source code of MULTINEST. This work is supported by
NSF AST-1109354, NSF AST-0908334 and NASA NNXI0AJ956.

\appendix
\section{The Likelihood Function}
\label{sec_like}

As discussed in the main text, the likelihood function describes 
the probability of the data given the model and its parameters, 
and for the present problem it is impossible to get a rigorous 
likelihood function. If, for
simplicity, the sampling uncertainty is assumed to be a Poisson
process, then the total variance from observations can be written as $
\sigma_{\rm obs}^{2} = \sigma_{\rm sys}^{2} + \left(\frac{\Phi_{\rm
   obs}}{n_{\rm obs}} \right)^{2}n_{\rm obs}$, where $n_{\rm obs}$ is
the observed number counts.  Replacing the Poisson process in the data
by that in the model, it can be shown that the variance in the
likelihood function can be approximated by $ \sigma_{\rm mod}^{2} =
\sigma_{\rm sys}^{2} + \left(\frac{\Phi_{\rm obs}}{n_{\rm obs}}
\right)^{2}\nu$, where $\nu$ is the number counts predicted by the
model. The likelihood for each stellar mass bin is then
\begin{eqnarray}
 \ln(L) & = & -\frac{1}{2} \frac{\left(\Phi_{\rm obs}-\Phi_{\rm mod}\right)^2}
               {\sigma_{\rm sys}^2 + \left(\frac{\Phi_{\rm obs}}{n_{\rm obs}} \right)^{2}\nu} \\ \nonumber 
        &   & -\frac{1}{2} \ln\left[2\pi\left(\sigma_{\rm sys}^2 
              + \left(\frac{\Phi_{\rm obs}}{n_{\rm obs}} \right)^{2}\nu\right)\right] \\ \nonumber 
        & = & -\frac{1}{2} \frac{\left(\Phi_{\rm obs}-\Phi_{\rm mod}\right)^2}  
             {\sigma_{\rm obs}^2 - \frac{\Phi_{\rm obs}}{n_{\rm obs}}\left(\Phi_{\rm obs} - \Phi_{\rm mod}\right)} \\ \nonumber
        &   & -\frac{1}{2} \ln\left[2\pi\left(\sigma_{\rm obs}^2 - \frac{\Phi_{\rm obs}}{n_{\rm obs}}\left(\Phi_{\rm obs} - \Phi_{\rm mod}\right)\right)\right]\,.
\end{eqnarray}
The second term in the variance, $\frac{\Phi_{\rm obs}}{n_{\rm obs}}
\left(\Phi_{\rm obs} - \Phi_{\rm mod}\right)$, can be evaluated if
$n_{\rm obs}$ is known.  In theory, this term makes the likelihood
deviate from a normal distribution, especially in the tails where
$\Phi_{\rm mod}$ is far from $\Phi_{\rm obs}$.  
We perform a series of tests and study how this term affects 
parameter estimation and
the computed value of the marginalised likelihood.  We find that for the
problems we study here the marginalised likelihood is hardly affected
because the posteriors for our models are always dominated by the
likelihood regions where $\Phi_{\rm mod}$ is close to $\Phi_{\rm obs}$.  
In this case the likelihood function reduces to the form given in  
Equation~(\ref{like}).

\section{Multinest sampling of the posterior distribution} 
\label{sec_multinest} 
 
The main goal of nested sampling is to evaluate the Bayesian evidence,
and to provide samples of the posterior distribution.  Briefly, the
algorithm works as follows. At the beginning of the process, one
randomly draws $N$ points in parameter space from the adopted
prior distribution. These points are called the active set. Each of
the points has a likelihood value $L_{i}$ ($i = 0, ..., N-1$), and
associated with it is an isolikelihood surface defined by the value of
$L_i$.  The volume (modulated by the prior distribution) 
within the surface is $X_{i}$.  The point
with the lowest likelihood value is denoted by $L^{(0)}$ and the
corresponding prior volume, $X^{(0)}$, can be approximated by the
total volume of the prior space. This point is removed from the active
set and is added to another list called the inactive set. A new point
with likelihood bigger than $L^{(0)}$ is then drawn from the prior
distribution and is added to the active set.  Before going to the next
iteration, it is important to know the volume (again modulated
by the prior distribution) occupied by the new
active set. Note that the ratios, $t_{i} \equiv X_{i}/X^{(0)}$ ($i =
0, ..., N-1$), can be considered as $N$ random numbers drawn from the
uniform distribution within $[0, 1)$. 

Define $t^{(1)} \equiv \max({t_{i}})$ and the
corresponding likelihood is $L^{(\rm 1)}$. Then the volume occupied
by the new active set is simply $X^{(\rm 1)} = t^{(\rm 1)}X^{(\rm
   0)}$.  The exact value of $t^{(\rm 1)}$ is unknown but it must
 satisfy the following distribution:
\begin{equation}
\label{nest_t}
 p(t) = Nt^{N-1},
\end{equation}
with the expectation of $\ln(t)$ equal to $-1/N$ and the standard
deviation in $\ln(t)$ equal to $1/N$.  Thus $t^{(\rm 1)}$ may be
approximated by the expectation value, $\exp(-1/N)$. The first step,
therefore, ends with a new active set that occupies a total volume
that is smaller by a factor of $t^{(\rm 1)}\approx \exp(-1/N)$ than
the old set, and with a new member, $\left(L^{(\rm 0)}, X^{(\rm
 0)}\right)$, in the inactive set.

By repeating the above process for the new active set produced at 
each subsequent step,  a list of points are drawn from the posterior
distribution:
\begin{equation}
 \{(L^{(k)}, X^{(k)})\}~~~~~~ \, \text{with} ~~\, X^{(k+1)} = t^{(k+1)}X^{(k)},
\end{equation}
which defines a series of nested shells in the parameter space and can
be used to sample the posterior distribution.  As described above, the
exact value of $t^{(k+1)}$ is unknown but can be approximated by
$\exp(-1/N)$ and the uncertainty can also be quantified with the use
of Equation~(\ref{nest_t}).  The Bayesian evidence is simply
\begin{equation} 
 Z = \sum_{k} L^{(k)}{[X^{( k+1)}-X^{( k-1)}] \over 2}\,.
\end{equation}
One iterates until the value of $Z$ reaches
a chosen accuracy, and we choose it to be $0.5$ in natural logarithmic
scale.

The efficiency of this algorithm depends on how efficiently the active
set can be replenished at each iteration.  Drawing new samples blindly
from the prior leads to a lower and lower acceptance rate as the
iso-likelihood surface shrinks.  However, if the surface can be
approximated by some regular shapes, then the active set can be
efficiently replenished.  The MULTINEST package developed by
\citet{Feroz09} provides such a method.  At each
iteration, multiple ellipsoids are used to approximate the
iso-likelihood surface of the new point drawn.  An ellipsoid can
either overlap with others or be isolated.  Too few big ellipsoids
would result in a bad approximation, while too many small ellipsoids
would result in too much overlap. In both cases the acceptance rate
would be low.  Optimal ellipsoidal decomposition is found by
minimising
\begin{equation}
 F = {\sum_{j} V(E_{j})\over V(S)} \geq 1\,,
\end{equation}
where $V(S)$ is the volume within the iso-likelihood surface and
$\sum_{j} V(E_{j})$ is the total volume of all the ellipsoids.  The
acceptance rate is simply the inverse of $F$.

The parameter that controls the process of posterior 
exploration is the size of the active set. A large active set can 
slow down the speed of going uphill  because after each iteration 
the size of the volume enclosed by the iso-likelihood surface 
shrinks by $\exp(-1/N)$. Conversely, to detect all the modes 
that are statistically significant in a high dimensional parameter space 
the size of the active set cannot be too small.
Unfortunately, there is no good way to find the optimal active set size. 
For this work, we use $2,000$ active points for Model I, Model II, and 
Model IIb, $5,000$ for Model III and $10,000$ for Model IV.  We
arrived at these values empirically by increasing the
active set size until the value of the marginal likelihood did not change
appreciably.

\label{lastpage}
\end{document}